\documentclass[aps, prd, letterpaper, 12pt, nofootinbib, superscriptaddress,longbibliography]{revtex4-2}
\usepackage[utf8]{inputenc}
\usepackage{amsmath,amssymb,amsfonts}
\usepackage{mathrsfs}
\usepackage{color}
\usepackage{graphicx} 
\usepackage[colorlinks=true,linkcolor=blue,citecolor=blue]{hyperref}

\allowdisplaybreaks

\begin{document}

%%%%%%%%%%%%%%%%%%%%%%%%%%%%%%%%%
\title{Loopy Determinations of $V_{ub}$ and $V_{cb}$}
%%%%%%%%%%%%%%%%%%%%%%%%%%%%%%%%%

\author{Wolfgang~Altmannshofer}
\email{waltmann@ucsc.edu}
\affiliation{Department of Physics, University of California Santa Cruz, and
Santa Cruz Institute for Particle Physics, 1156 High St., Santa Cruz, CA 95064, USA}

\author{Nathan~Lewis}
\email{nathan.lewis@sjsu.edu}
\affiliation{ Department of Physics, San Jose State University, 1 Washington Sq, San Jose, CA 95192, USA}

\begin{abstract}
We use loop induced processes like meson oscillations and rare $b$ hadron decays to determine the absolute values of the CKM matrix elements $|V_{ub}|$ and $|V_{cb}|$ and compare our results to the standard determinations based on inclusive and exclusive semileptonic tree-level decays of $B$ mesons. For many years there have been tensions between the inclusive and exclusive determinations. Assuming the absence of new physics, we find that meson oscillation data shows a slight preference for the inclusive value of $|V_{cb}|$ and the exclusive value for $|V_{ub}|$. Rare $b$ decay data prefers values for $|V_{cb}|$ far below the inclusive and exclusive determinations, offering a new perspective on some of the persistent rare $b$ decay anomalies. 
\end{abstract}

\maketitle

%%%%%%%%%%%%%%%%%%%%%%%%%%%%%%%%%
\section{Introduction}
%%%%%%%%%%%%%%%%%%%%%%%%%%%%%%%%%

The elements $V_{ub}$ and $V_{cb}$ of the Cabibbo-Kobayashi-Maskawa (CKM) matrix~\cite{Cabibbo:1963yz,Kobayashi:1973fv} are crucial input parameters for the theory predictions of many observables in the flavor sector. A standard strategy to determine the absolute values of $V_{ub}$ and $V_{cb}$ is to use tree level $b \to c \ell \nu$ and $b \to u \ell \nu$ decays, assuming no new physics at tree level. 
These values are then used to make Standard Model (SM) predictions for loop level flavor changing neutral current processes that are highly sensitive to new physics effects.

Since many years, discrepancies exist between the tree level determinations using the inclusive $B \to X_c \ell \nu$ and $B \to X_u \ell \nu$ decays on the one side, and exclusive decays, like $B \to D^{(*)} \ell \nu$ and $B \to \pi \ell \nu$ on the other~\cite{Ricciardi:2019zph, Ricciardi:2021shl, Zyla:2020zbs} (also $B \to \pi \pi \ell \nu$ decays can be used to determine $|V_{ub}|$~\cite{Kang:2013jaa}). These discrepancies limit the sensitivity of various loop processes to new physics. At the same time, explaining the discrepancies by new physics is challenging~\cite{Crivellin:2014zpa, Bernlochner:2014ova, Colangelo:2016ymy}.
Improving the theoretical description of inclusive and exclusive $b \to c \ell \nu$ and $b \to u \ell \nu$ decays is thus of high priority (see~\cite{Gambino:2020crt, Ferlewicz:2020lxm, Fael:2020njb, Fael:2020tow, Capdevila:2021vkf, Leljak:2021vte, Biswas:2021qyq, Mannel:2021mwe, Martinelli:2021onb, FermilabLattice:2021cdg, Bordone:2021oof, FermilabLattice:2021bxu, Gonzalez-Solis:2021awb, Bansal:2021oon} for recent progress). Alternatively, one can focus on combinations of loop observables that are independent of $V_{cb}$ and/or $V_{ub}$ as emphasized for example in~\cite{Buras:2003td, Bobeth:2021cxm, Buras:2021nns}. 

In this paper, we follow a different approach. Assuming that there is no new physics in loop induced flavor changing neutral current processes, we use those processes to determine $|V_{ub}|$ and $|V_{cb}|$ and compare them to the tree-level determinations. The main results of our study are loop level determinations of $|V_{ub}|$ and $|V_{cb}|$ based on meson oscillations and determinations of $|V_{cb}|$ from rare $b$ hadron decays.

On the one hand, we are motivated by recent progress in the SM calculations of the parameter $\epsilon_K$ that quantifies indirect CP violation in neutral Kaon oscillations~\cite{Brod:2019rzc,Brod:2021qvc}. As $|V_{cb}|$ is the largest uncertainty of the SM prediction, the precisely known experimental value of $\epsilon_K$ gives one of the most precise determinations of $|V_{cb}|$.
On the other hand, the observation of anomalously low branching ratios of the rare $B$ meson decays $B \to K \mu^+ \mu^-$, $B \to K^* \mu^+ \mu^-$, and $B_s \to \phi \mu^+ \mu^-$ by the LHCb collaboration~\cite{Aaij:2014pli, Aaij:2016flj, LHCb:2021zwz} provides motivation to re-examine the possible role of $|V_{cb}|$ in these $B$ anomalies (see e.g.~\cite{Altmannshofer:2014rta} for a previous study).

We stress that our results for $|V_{ub}|$ and $|V_{cb}|$ should not be used in studies of new physics in flavor changing processes. Instead, they offer a new perspective on the tension between inclusive and exclusive $|V_{ub}|$ and $|V_{cb}|$ determinations, the role of $\epsilon_K$ in CKM fits, and some of the persistent anomalies in rare $B$ meson decays.   

This paper is organized as follows: In section~\ref{sec:direct} we review $|V_{ub}|$ and $|V_{cb}|$ determinations from tree level $b$ hadron decays, establishing a baseline for the comparison with our loop level determinations. In section~\ref{sec:loop_obs} we discuss a number of loop observables that are sensitive to $|V_{ub}|$ and $|V_{cb}|$. In particular, we consider neutral Kaon and $B$ meson mixing as well as rare decays of $b$ hadrons. Our results of the loop level determinations are discussed in section~\ref{sec:fits}. We conclude in section~\ref{sec:conclusions}.

%%%%%%%%%%%%%%%%%%%%%%%%%%%%%%%%%%%%%%%%%%%%%%%%%%
\section{Tree Level Determinations of \texorpdfstring{$V_{cb}$}{Vcb} and \texorpdfstring{$V_{ub}$}{Vub}} \label{sec:direct}
%%%%%%%%%%%%%%%%%%%%%%%%%%%%%%%%%%%%%%%%%%%%%%%%%%

Direct tree-level determinations of the CKM matrix elements $|V_{ub}|$ and $|V_{cb}|$ come from measurements of inclusive and exclusive semileptonic decays based on the $b \to u \ell \nu$ and $b \to c \ell \nu$ transitions. For many years, notable discrepancies have existed between inclusive and exclusive determinations. 

The Particle Data Group (PDG) quotes the following inclusive values~\cite{Zyla:2020zbs}
\begin{equation} \label{eq:VcbVub_incl}
    |V_{cb}|_\text{incl.} = (42.2 \pm 0.8)\times 10^{-3}~, \qquad |V_{ub}|_\text{incl.} = (4.25 \pm 0.30) \times 10^{-3}~,
\end{equation}
where we added all individual uncertainties in quadrature.
Note that the above value for $|V_{ub}|_\text{incl.}$ does not yet take into account the latest determination from the Belle collaboration, $|V_{ub}|_\text{incl.} = (4.1 \pm 0.28) \times 10^{-3}$ which is compatible within the uncertainties but has a slightly lower central value~\cite{Belle:2021eni}. Also note that the PDG value for $|V_{cb}|_\text{incl.}$ does not yet take into account improved theory calculations which give $|V_{cb}|_\text{incl.} = (42.16 \pm 0.51) \times 10^{-3}$~\cite{Bordone:2021oof}. For definiteness we will use the PDG values in \eqref{eq:VcbVub_incl} for the remainder of this work.

For exclusive determinations, the PDG gives averages based on $\bar B \to D \ell \nu$ and $\bar B \to D^* \ell \nu$ decays for $|V_{cb}|$ and $\bar B \to \pi \ell \nu$ decays for $|V_{ub}|$. The averages read~\cite{Zyla:2020zbs}
\begin{equation} \label{eq:VcbVub_excl}
    |V_{cb}|_\text{excl.} = (39.5 \pm 0.9)\times 10^{-3}~, \qquad |V_{ub}|_\text{excl.} = (3.70 \pm 0.16) \times 10^{-3}~,
\end{equation}
where, as above, we added all individual uncertainties in quadrature.
We note that the PDG average of $|V_{cb}|_\text{excl.}$ does not yet include the latest lattice results on the $B \to D^*$ form factors from~\cite{FermilabLattice:2021cdg}. The value for $|V_{cb}|$ determined in~\cite{FermilabLattice:2021cdg} is $|V_{cb}|_{B\to D^*} = (38.4 \pm 0.74)\times 10^{-3}$, very similar to the results from~\cite{Belle:2018ezy} and \cite{BaBar:2019vpl} that enter the PDG average, but with slightly smaller uncertainties. The discrepancies between the inclusive and exclusive values quoted by the PDG are $2.4\sigma$ for $V_{cb}$ and $1.6\sigma$ for $V_{ub}$\,.

Exclusive decays are also used to independently determine the ratio $|V_{ub}|/|V_{cb}|$.
A measurement from LHCb~\cite{Aaij:2015bfa} of the branching ratios $\Lambda_b^0 \to p \mu^- \nu$ and $\Lambda_b^0 \to \Lambda_c^+ \mu^- \nu$, combined with form factor ratios from~\cite{Detmold:2015aaa}, gives~\cite{Amhis:2019ckw}
\begin{equation} \label{eq:VcbVub_Lambda} 
|V_{ub}|/|V_{cb}| \Big|_{\Lambda_b} = 0.079 \pm 0.006~.
\end{equation}
Combining recent results from LHCb on the decay modes $B_s \to K^- \mu^+ \nu$ and $B_s \to D_s^- \mu^+ \nu$~\cite{LHCb:2020ist} with lattice form factors from~\cite{Bouchard:2014ypa, Flynn:2015mha, FermilabLattice:2019ikx, McLean:2019qcx} gives~\cite{Aoki:2021kgd}
\begin{equation} \label{eq:VubVcb_Bs_lattice}
|V_{ub}|/|V_{cb}| \Big|_{B_s} = \begin{cases} 0.0819 \pm 0.0077 ~, \quad \text{for}~ q^2 < 7~\text{GeV}^2 ~, \\ 0.0860 \pm 0.0053 ~, \quad \text{for}~ q^2 > 7~\text{GeV}^2 ~. \end{cases}
\end{equation}
The values for the ratios show good agreement between the high $q^2$ and low $q^2$ region and with the values obtained from $\Lambda_b$ decays quoted above.
If alternative form factors based on light cone sum rules~\cite{Khodjamirian:2017fxg} are used to determine $|V_{ub}|/|V_{cb}| \Big|_{B_s}$ at low $q^2$, one finds a significantly lower value. We will work with the lattice results~\eqref{eq:VubVcb_Bs_lattice} in the following.

Finally, the exclusive leptonic decay mode $B^+ \to \tau^+ \nu$ can also be used for a tree level determination of $|V_{ub}|$. In the SM, the $B^+ \to \tau^+ \nu$ branching ratio is given by
\begin{equation}
    \text{BR}(B^+ \to \tau^+ \nu)_\text{SM} = \tau_{B^+} \frac{G_F^2}{8\pi} f_{B^+}^2 m_{B^+} m_\tau^2 \left(1 - \frac{m_\tau^2}{m_{B^+}^2}\right)^2 |V_{ub}|^2 ~.
\end{equation}
The Fermi constant, the tau mass, and the $B^+$ mass and lifetime have negligible uncertainties~\cite{Zyla:2020zbs}. The dominant uncertainty comes from the $B^+$ meson decay constant. For our numerical analysis, we use the value given by the flavor lattice averaging group (FLAG), $f_{B^+} \simeq f_{B} = (190.0 \pm 1.3)$\,MeV~\cite{Aoki:2021kgd}, ignoring small isospin breaking effects.

The experimental world average of the $B^+ \to \tau^+ \nu$ branching ratio quoted by the heavy flavor averaging group (HFLAV)~\cite{Amhis:2019ckw} is based on results by BaBar~\cite{BaBar:2012nus} and Belle~\cite{Belle:2015odw} and reads
\begin{equation}
    \text{BR}(B^+ \to \tau^+ \nu)_\text{exp} = (1.06 \pm 0.19) \times 10^{-4} ~.
\end{equation}
This allows us to extract the following value for $|V_{ub}|$
\begin{equation} \label{eq:Vub_Btaunu}
    |V_{ub}|_{B \to \tau\nu} = (4.08 \pm 0.37 \pm 0.03) \times 10^{-3}~,
\end{equation}
where the first uncertainty is experimental and the second theoretical (i.e. from the $B^+$ meson decay constant). Our central value agrees very well with the values quoted in~\cite{Aoki:2021kgd}.

%%%%%%%%%%%%%%%%%%%%%%%%%%%%%%%%%%%%%%%%
\begin{figure}[tb]
\centering
\includegraphics[width=0.6\textwidth]{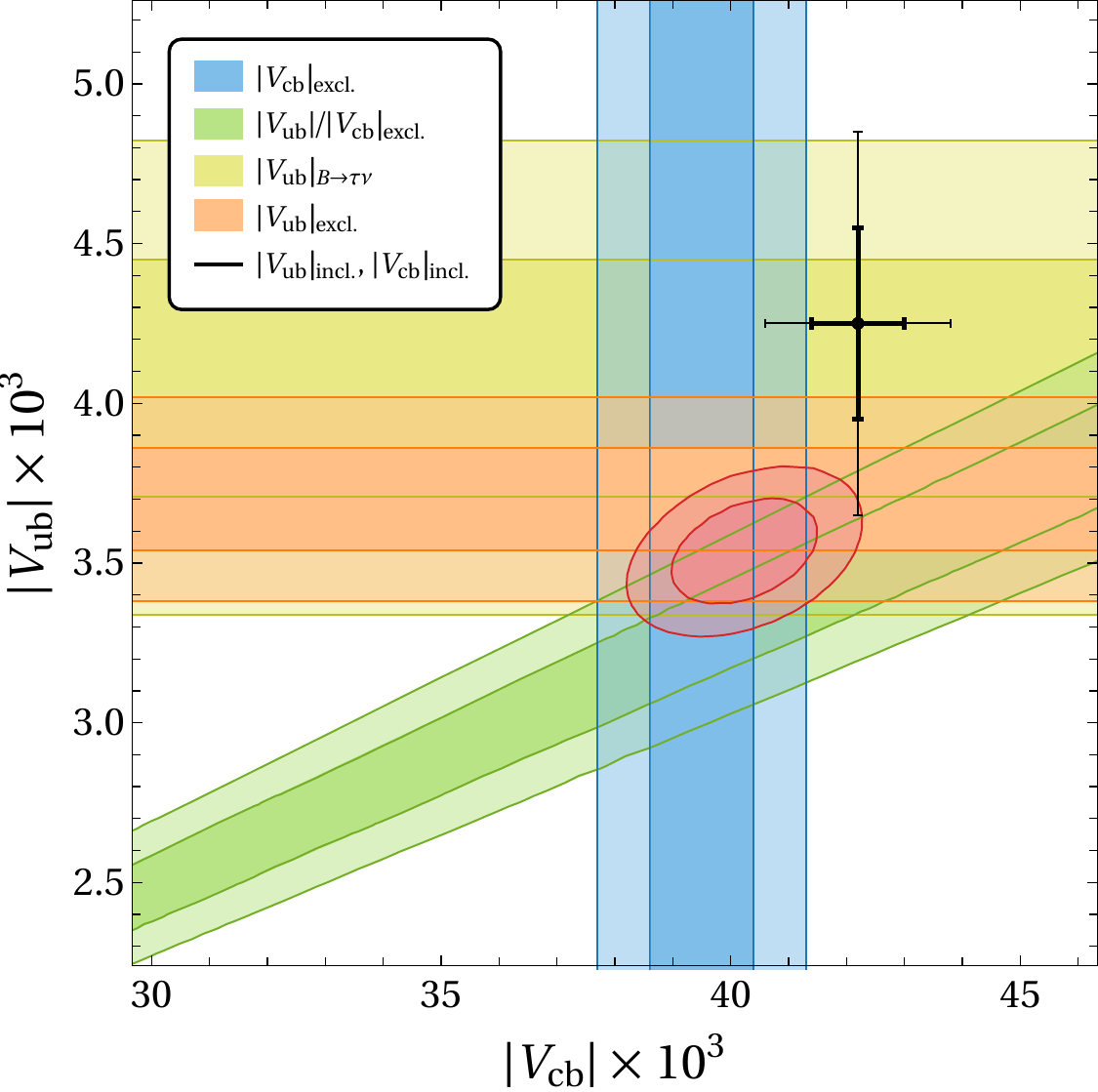} 
\caption{Comparison of inclusive and exclusive determinations of the CKM matrix elements $|V_{cb}|$ and $|V_{ub}|$. The black error bars show the PDG average of the inclusive values at $1\sigma$ and $2\sigma$, while the red region shows our combination of all exclusive determinations at $1\sigma$ and $2\sigma$.}
\label{fig:incl_excl}
\end{figure}
%%%%%%%%%%%%%%%%%%%%%%%%%%%%%%%%%%%%%%%%

In Figure~\ref{fig:incl_excl}, we compare the various determinations in the $|V_{cb}| - |V_{ub}|$ plane. The black error bars correspond to the $1\sigma$ and $2\sigma$ ranges for the inclusive values given in~\eqref{eq:VcbVub_incl}. The $1\sigma$ and $2\sigma$ ranges of the exclusive values of $V_{cb}$ and $V_{ub}$ in~\eqref{eq:VcbVub_excl} are represented by the blue and orange bands, respectively. The green band shows the combination of the ratios in~\eqref{eq:VcbVub_Lambda} and~\eqref{eq:VubVcb_Bs_lattice}, while the yellow band shows the result from the $B^+ \to \tau^+ \nu$ decay in~\eqref{eq:Vub_Btaunu}. 
Finally, the red region corresponds to the combination of all exclusive determinations. Approximating this region by a 2-dimensional Gaussian distribution, we find
\begin{equation}
    |V_{cb}|_\text{excl.} = (40.24 \pm 0.83)\times 10^{-3}~, \qquad |V_{ub}|_\text{excl.} = (3.54 \pm 0.11) \times 10^{-3}~.
\end{equation}
with an error correlation of $\rho = +37\%$\,. We note that our results for $|V_{cb}|_\text{excl.}$ and $|V_{ub}|_\text{excl.}$ are very close to the results of the flavor lattice averaging group (FLAG)~\cite{Aoki:2021kgd}. Our value for $|V_{cb}|_\text{excl.}$ is slightly larger and has a $\sim 20\%$ larger uncertainty than the FLAG result ($|V_{cb}|_\text{excl.}^\text{FLAG} = (39.48 \pm 0.68)\times 10^{-3}$~\cite{Aoki:2021kgd}). This is due to the fact that we use as input the more conservative PDG value from~\eqref{eq:VcbVub_excl}.

The results shown in Figure~\ref{fig:incl_excl} serve as a baseline for the comparison with alternative determinations of $|V_{cb}|$ and $|V_{ub}|$, which are discussed in the following sections.

%%%%%%%%%%%%%%%%%%%%%%%%%%%%%%%%%
\section{Loop Observables Sensitive to \texorpdfstring{$V_{cb}$}{Vcb} and \texorpdfstring{$V_{ub}$}{Vub}} \label{sec:loop_obs}
%%%%%%%%%%%%%%%%%%%%%%%%%%%%%%%%%

While it is customary to work with the Wolfenstein parameterization of the CKM matrix~\cite{Wolfenstein:1983yz}, we decide to directly parameterize the CKM matrix in terms of the sine of the Cabibbo angle $\lambda = \sin\theta_C$, the absolute values $|V_{cb}|$ and $|V_{ub}|$, and the phase $\gamma = \text{arg}(-V_{ud}V_{ub}^*/V_{cd}V_{cb}^*)$. In this way, the $|V_{cb}|$ and $|V_{ub}|$ dependence of the various observables discussed below is made fully transparent. 
Working in the standard phase convention for the CKM matrix and expanding in $\lambda$, we have
\begin{align} \label{eq:CKM}
V_{ud} &\simeq 1 - \frac{\lambda^2}{2} ~,& V_{us} &\simeq \lambda ~, &  V_{ub} &\simeq |V_{ub}| e^{-i \gamma} ~, \nonumber \\
V_{cd} &\simeq - \lambda ~,& V_{cs} &\simeq 1 - \frac{\lambda^2}{2} ~, & V_{cb} &= |V_{cb}| ~, \nonumber \\
V_{td} &\simeq |V_{cb}|\lambda - |V_{ub}|e^{i \gamma} \left(1-\frac{\lambda^2}{2}\right)~,& V_{ts} &\simeq -|V_{cb}|\left(1-\frac{\lambda^2}{2}\right) - |V_{ub}|\lambda e^{i \gamma} ~, &  V_{tb} &\simeq 1 ~,
\end{align}
where for each CKM element we take into account relative corrections up to $\mathcal O(\lambda^2)$.

%%%%%%%%%%%%%%%%%%%%%%%%%%%%%%%%%%%%%%%%%%%%%%%%%%
\subsection{CP violation in neutral kaon mixing} \label{sec:epsK}
%%%%%%%%%%%%%%%%%%%%%%%%%%%%%%%%%%%%%%%%%%%%%%%%%%

The parameter $\epsilon_K$ is a measure of CP violation in neutral kaon oscillations. Its SM prediction depends strongly on the value of $|V_{cb}|$. It has been shown in~\cite{Brod:2019rzc} that the short distance theory uncertainty of $\epsilon_K$ can be strongly reduced through a reorganization of the relevant effective Hamiltonian. The uncertainty of the SM prediction of $\epsilon_K$ is then dominated by the CKM matrix input, thus strengthening the role of $\epsilon_K$ in determining CKM parameters. Using the expressions from~\cite{Brod:2019rzc}, we can write 
\begin{equation} \label{eq:epsK_raw}
    |\epsilon_K|_\text{SM} = \kappa_\epsilon C_\epsilon \hat B_K \left( -\frac{1}{2} \text{Im}((V_{ts}^*V_{td})^2) \eta_{tt} \mathscr S(x_t) - \text{Im}(V_{us}^*V_{ud}V_{ts}^*V_{td}) \eta_{ut} \mathscr S(x_c,x_t) \right) ~,
\end{equation}
where $x_q = m_q^2/m_W^2$ is the ratio of quark mass and the $W$ boson mass squared. The loop functions include the leading terms in $x_c$ and are given by
\begin{align} \label{eq:S0}
\mathscr S(x_t) &= \frac{4x_t-11x_t^2+x_t^3}{4(1-x_t)^2}-\frac{3x_t^3\log x_t}{2(1-x_t)^3} ~, \\
\mathscr S(x_c,x_t) &= x_c\left( 1 - \log(x_t/x_c) + \frac{3 x_t}{4(1-x_t)}+\frac{3x_t^2\log x_t}{4(1-x_t)^2} \right) ~.
\end{align}
We follow~\cite{Brod:2021qvc} and include electroweak corrections of the top contribution by making the replacement $\eta_{tt} \to \eta_{tt}(1-\Delta_{tt})$. To obtain numerical predictions we use the input $\Delta_{tt} = 0.01 \pm 0.004$~\cite{Brod:2021qvc}, $\eta_{tt} = 0.55 \pm 0.02$~\cite{Brod:2019rzc}, $\eta_{ut} = 0.402 \pm 0.005$~\cite{Brod:2019rzc}, $m_t = (163.48 \pm 0.86)\,$GeV~\cite{Brod:2019rzc}, $m_c = (1.27 \pm 0.02)\,$GeV~\cite{Zyla:2020zbs}, $\kappa_\epsilon = 0.94 \pm 0.02$~\cite{Buras:2008nn}, and $\hat B_K = 0.7625 \pm 0.0097$~\cite{Aoki:2021kgd}. The coefficient $C_\epsilon$ is given by~\cite{Brod:2019rzc} 
\begin{equation}
    C_\epsilon = \frac{G_F^2 F_K^2 m_{K^0} m_W^2}{6 \sqrt{2} \pi^2 \Delta M_K} \simeq 3.634 \times 10^4  ~,
\end{equation}
with negligible uncertainty. Using these numerical values and expressing the CKM matrix elements in terms of $|V_{cb}|$, $|V_{ub}|$, $\lambda$, and $\gamma$ as in~\eqref{eq:CKM}, we find
\begin{multline} \label{eq:epsK}
    |\epsilon_K|_\text{SM} = \left[\left(|V_{cb}|^2\lambda\left(1-\frac{\lambda^2}{2}\right) - |V_{cb}||V_{ub}| (1-2\lambda^2)\cos\gamma - |V_{ub}|^2 \lambda \right) a_{\epsilon_K} \right. \\ \left. + \lambda \left(1-\frac{\lambda^2}{2}\right) b_{\epsilon_K} \right] |V_{cb}||V_{ub}|\sin\gamma~,
\end{multline}
with the coefficients
\begin{equation}
a_{\epsilon_K} = (3.29 \pm 0.15)\times 10^4 ~,\qquad b_{\epsilon_K} = 20.55\pm 0.81 ~.
\end{equation}
For the errors of the numerical coefficients $a_{\epsilon_K}$ and $b_{\epsilon_K}$, we find a correlation of $+34\%$, which we take into account in our numerical analysis in section~\ref{sec:fits}. 

Experimentally, $\epsilon_K$ is known with very high precision~\cite{Zyla:2020zbs}
\begin{equation}
|\epsilon_K|_\text{exp} = (2.228 \pm 0.011 )\times 10^{-3}~.
\end{equation}
The measurement provides a stringent constraint on the CKM parameters in~(\ref{eq:epsK}).

%%%%%%%%%%%%%%%%%%%%%%%%%%%%%%%%%%%%%%%%%%%%%%%%%%
\subsection{Neutral B meson mixing} \label{sec:Bmixing}
%%%%%%%%%%%%%%%%%%%%%%%%%%%%%%%%%%%%%%%%%%%%%%%%%%

CP violation in $B^0 - \bar B^0$ mixing provides a very important ingredient for fits of the CKM matrix. In particular, the CKM angle $\beta = \text{Arg}(- V_{cd} V_{cb}^*/V_{td}V_{tb}^*)$ can be accessed through measurements of time-dependent CP asymmetries in $b \to c \bar c s$ transitions.
The world average from HFLAV is~\cite{Amhis:2019ckw} 
\begin{equation}
    \sin(2\beta)_\text{exp.} = 0.699 \pm 0.017 ~.
\end{equation}
Expressing $\sin(2\beta)$ in terms of $\lambda$, $|V_{cb}|$, $|V_{ub}|$ and $\gamma$, we find the SM prediction
\begin{multline}
        \sin(2\beta)_\text{SM} =\frac{2|V_{ub}| (|V_{cb}|\lambda - |V_{ub}| \cos\gamma )\sin\gamma }{( |V_{ub}| \cos\gamma - |V_{cb}|\lambda )^2 + |V_{ub}|^2\sin^2\gamma} \\ 
        \times \left( 1 + \frac{|V_{cb}|\lambda^3 }{2(|V_{ub}| \cos\gamma-|V_{cb}|\lambda )} \frac{( |V_{ub}| \cos\gamma - |V_{cb}|\lambda )^2 - |V_{ub}|^2\sin^2\gamma}{( |V_{ub}| \cos\gamma - |V_{cb}|\lambda )^2 + |V_{ub}|^2\sin^2\gamma} \right) ~,
\end{multline}
where we have neglected terms of $\mathcal O(\lambda^4)$.

Also the neutral $B$ meson oscillation frequencies $\Delta M_d$ and $\Delta M_s$ can be used to determine CKM parameters. From the experimental side, the frequencies are known with remarkable precision 
\begin{equation}
    \Delta M_d^\text{exp.} = (0.5065 \pm 0.0019)\, \text{ps}^{-1} ~,\qquad \Delta M_s^\text{exp.} = (17.7656 \pm 0.0056)\, \text{ps}^{-1} ~,
\end{equation}
where we quote the experimental world average for $\Delta M_d^\text{exp.}$ from HFLAV~\cite{Amhis:2019ckw} and   the recent measurement for $\Delta M_s^\text{exp.}$ by LHCb~\cite{LHCb:2021moh}, which is the single most precise determination to date.

The SM predictions of the oscillation frequencies are given by the well known expressions
\begin{eqnarray}
\Delta M_d^\text{SM} &=& \frac{G_F^2 m_W^2}{6 \pi^2} m_{B_d} |V_{td}^* V_{tb}|^2 S_0(m_t^2/m_W^2) \eta_B f_{B_d}^2 \hat B_{B_d} ~,\\
\Delta M_s^\text{SM} &=& \frac{G_F^2 m_W^2}{6 \pi^2} m_{B_s} |V_{ts}^* V_{tb}|^2 S_0(m_t^2/m_W^2) \eta_B f_{B_s}^2 \hat B_{B_s} ~.
\end{eqnarray}
The dominant uncertainty stems from the values of the hadronic matrix elements, parameterized by the $B$ meson decay constants $f_{B_q}$ and the so-called bag parameters $\hat B_{B_q}$. We use the latest $N_f = 2+1+1$ lattice results from~\cite{Dowdall:2019bea}, $f_{B_s} \sqrt{\hat B_{B_s}} = (256.1 \pm 5.7)$\,MeV, $f_{B_d} \sqrt{\hat B_{B_d}} = (210.6 \pm 5.5) $\,MeV, and $\xi = f_{B_s} \sqrt{\hat B_{B_s}} / f_{B_d} \sqrt{\hat B_{B_d}} = 1.216 \pm 0.016$, neglecting unknown correlations between the uncertainties. The quoted lattice results for the hadronic matrix elements agree well with alternative determinations using sum rules~\cite{King:2019lal}. (Note that the FLAG averages of the older $N_f = 2+1$ hadronic matrix elements~\cite{Aoki:2021kgd} are instead significantly larger.) In our theory predictions of the mass differences, we also take into account the uncertainty from the top mass $m_t = (163.48 \pm 0.86)$\,GeV~\cite{Brod:2019rzc} that enters the loop function $S_0(x) = \mathscr S(x)$ given in~\eqref{eq:S0}.
We do not take into account the uncertainty in the higher order QCD correction factor $\eta_B \simeq 0.552$~\cite{DiLuzio:2017fdq}, as it is negligibly small. Using PDG values for $G_F$, $m_W$, and $m_{B_q}$~\cite{Zyla:2020zbs}, we find the expressions
\begin{eqnarray}
    \Delta M_d^\text{SM} &=& \left( \lambda^2 |V_{cb}|^2 + |V_{ub}|^2(1-\lambda^2) - \lambda |V_{cb}||V_{ub}|(2 - \lambda^2) \cos\gamma  \right) a_{\Delta M_d}~, \\
\Delta M_s^\text{SM} &=& \left( |V_{cb}|^2(1-\lambda^2) + 2 \lambda |V_{cb}||V_{ub}| \cos\gamma \right) a_{\Delta M_s} ~,
\end{eqnarray}
with the numerical coefficients
\begin{equation}
    a_{\Delta M_d} = ( 6.77 \pm 0.26)\times 10^3 \,\text{ps}^{-1}  ~,\qquad 
    a_{\Delta M_s} = ( 10.18 \pm 0.37)\times 10^3 \,\text{ps}^{-1} ~.
\end{equation}
We find a sizable error correlation of $\rho = +76\%$ between the two coefficients. The correlation arises from the precisely known ratio $\xi$.

Note that the dependence of the meson mixing observables on the top mass also leads in principle to a correlation between the coefficients $a_{\Delta M_q}$, which are relevant for $B$ mixing, and the coefficients $a_{\epsilon_K}$, $b_{\epsilon_K}$, which enter the expression for $\epsilon_K^\text{SM}$ discussed above.
However, since the top mass uncertainty is not significant in both $\epsilon_K^\text{SM}$ and the oscillation frequencies $\Delta M_q^\text{SM}$, it is justified to neglect the error correlation across these observables.

%%%%%%%%%%%%%%%%%%%%%%%%%%%%%%%%%%%%%%%%%%%%%%%%%%
\subsection{Rare b hadron decays} \label{sec:rareB}
%%%%%%%%%%%%%%%%%%%%%%%%%%%%%%%%%%%%%%%%%%%%%%%%%%

The branching ratios of flavor changing neutral current $b$ hadron decays depend on the CKM matrix element combinations $|V_{ts}^* V_{tb}|^2$ or $|V_{td}^* V_{tb}|^2$ and can be used to constrain $|V_{cb}|$ and, to a lesser extent, $|V_{ub}|$. In the following, we consider the radiative decay $B \to X_s \gamma$, the leptonic decays $B_s \to \mu^+ \mu^-$ and $B^0 \to \mu^+\mu^-$, and the semi-leptonic decays $B^+ \to K^+ \mu^+\mu^-$, $B^0 \to K^{*\,0} \mu^+\mu^-$, $B_s \to \phi \mu^+\mu^-$, and $\Lambda_b \to \Lambda \mu^+\mu^-$. We do not include additional decay modes like $B^0 \to K^0 \mu^+ \mu^-$ or $B^+ \to K^{*\,+} \mu^+\mu^-$, as they have larger experimental uncertainties and therefore have little impact on our results.

The inclusive radiative decay $B \to X_s \gamma$ is known to be a very sensitive probe of new physics. In the absence of new physics, it can be used to determine the CKM matrix element $V_{cb}$. The $B \to X_s \gamma$ rate is to a large extent determined by the product of the Wilson coefficient $|C_7^\text{incl}|$ and the CKM matrix combination $|V_{tb} V_{ts}^*|$. A recent global fit to inclusive $B \to X_s \gamma$ measurements finds~\cite{Bernlochner:2020jlt}
\begin{equation} \label{eq:C7exp}
    |C_7^\text{incl} \,V_{tb} V_{ts}^*|  = (14.77 \pm 0.78)\times 10^{-3} ~,
\end{equation}
which can be compared to the SM prediction~\cite{Bernlochner:2020jlt}
\begin{equation} \label{eq:C7SM}
    |C_7^\text{incl} \,V_{tb} V_{ts}^*|_\text{SM}  = \left(|V_{cb}| (1-\frac{\lambda^2}{2}) + |V_{ub}|\lambda \cos \gamma \right) \times (0.3624 \pm 0.0151) ~,
\end{equation}
where we added the uncertainties from $c \bar c$ loops and from scale variation in quadrature. Combining~\eqref{eq:C7exp} with~\eqref{eq:C7SM} allows one to constrain the value of $|V_{cb}|$. The term in~\eqref{eq:C7SM} proportional to $|V_{ub}|$ plays  a negligible role. 

The leptonic decays $B_s \to \mu^+\mu^-$ and $B^0 \to \mu^+\mu^-$ are of particular interest, as they are theoretically very clean, with the most relevant hadronic uncertainty coming from the $B_s$ and $B^0$ meson decay constants. The latest lattice results for decay constants~\cite{Bazavov:2017lyh} have reached sub-percent precision, implying that the dominant source of uncertainty in the SM predictions of the branching ratios BR$(B_s \to \mu^+\mu^-)$ and BR$(B^0 \to \mu^+\mu^-)$ is the CKM matrix input, in particular $|V_{cb}|$. Assuming the absence of new physics, precision measurements of BR$(B_s \to \mu^+\mu^-)$ and BR$(B^0 \to \mu^+\mu^-)$ can therefore give important constraints on $|V_{cb}|$. We use the experimental world average of the branching ratios that has been determined in~\cite{Altmannshofer:2021qrr}, based on the most recent experimental results on $B_s \to \mu^+\mu^-$ and $B^0 \to \mu^+\mu^-$ from the LHCb, ATLAS, and CMS collaborations~\cite{LHCb:2021awg, LHCb:2021vsc, Aaboud:2018mst, Sirunyan:2019xdu}
\begin{align}
\text{BR}(B_s \to \mu^+\mu^-)_\text{exp.} &= (2.93 \pm 0.35) \times 10^{-9} ~, \\
\text{BR}(B^0 \to \mu^+\mu^-)_\text{exp.} &= (0.56 \pm 0.70) \times 10^{-10} ~, 
\end{align}
with an error correlation of $\rho = -27\%$\,. 

The SM predictions for the branching ratios are proportional to the CKM factors $|V_{tb}V_{ts}^*|^2$ and $|V_{tb}V_{td}^*|^2$, respectively. Therefore, we can write the branching ratios as
\begin{eqnarray}
\text{BR}(B_s \to \mu^+\mu^-)_\text{SM} &=& \left(|V_{cb}|^2 (1-\lambda^2) + 2 \lambda |V_{cb}||V_{ub}| \cos \gamma \right) a_{B_s\to\mu\mu}~,  \\
\text{BR}(B^0 \to \mu^+\mu^-)_\text{SM} &=& \left( \lambda^2 |V_{cb}|^2 + |V_{ub}|^2(1-\lambda^2) - \lambda |V_{cb}||V_{ub}|(2 - \lambda^2) \cos\gamma  \right) a_{B^0\to\mu\mu}~.
\end{eqnarray}
We determine the numerical coefficients $a_{B_s\to \mu\mu}$ and $a_{B^0\to \mu\mu}$ following the SM calculation of the branching ratios in~\cite{Beneke:2019slt}. We use the averages of the $B$ meson decay constants from FLAG assuming isospin symmetry, $f_{B_s} = (230.3 \pm 1.3)$\,MeV, $f_{B} = (190.0 \pm 1.3)$\,MeV~\cite{Aoki:2021kgd}, as well as the very precisely known ratio $f_{B_s}/f_{B} = 1.209 \pm 0.005$~\cite{Aoki:2021kgd}. We also take into account the uncertainties from the $B$ meson lifetimes $\tau_{B_s^H} = (1.615 \pm 0.009)$\,ps, $\tau_{B^0} = (1.520 \pm 0.004)$\,ps~\cite{Zyla:2020zbs}, a relative uncertainty of $1.1\%$ from the top mass~\cite{Beneke:2019slt}, an uncertainty of $1.2\%$ from scale variation and higher order corrections~\cite{Beneke:2019slt}, and a $0.5\%$ uncertainty from light cone distribution amplitudes that enter the QED corrections of the branching rations~\cite{Beneke:2019slt}. In that way we obtain
\begin{equation}
    a_{B_s\to \mu\mu} = (2.15 \pm 0.04)\times 10^{-6} ~, \qquad a_{B^0\to \mu\mu} = (1.36 \pm 0.03)\times 10^{-6} ~,
\end{equation}
with a large positive error correlation of $\rho = +85\%$.

Similarly, semileptonic decays of $b$ hadrons can be used to determine $|V_{cb}|$, albeit with larger theoretical uncertainty.
The branching ratios of the semileptonic rare $b$ hadron decays are measured in bins of $q^2$, the di-lepton invariant mass squared. 
As in the case of the $B_s \to \mu^+\mu^-$ decay, we factor out the CKM dependence and write the SM predictions for the $b \to s \mu\mu$ branching ratios of a hadron $H_1$ to a hadron $H_2$ and two muons in the following way:
\begin{equation} \label{eq:semileptonicBR}
   \text{BR}(H_1 \to H_2 \mu^+\mu^-)_\text{SM}^{[q_\text{min}^2,q_\text{max}^2]} = \left(|V_{cb}|^2 (1-\lambda^2) + 2 \lambda |V_{cb}||V_{ub}| \cos \gamma \right) a_{H_1 \to H_2 \mu\mu}^{[q_\text{min}^2,q_\text{max}^2]} ~,
\end{equation}
where the superscript $[q_\text{min}^2,q_\text{max}^2]$ indicates the $q^2$ bin. In our analysis, we consider the well measured $B^+ \to K^+ \mu^+\mu^-$, $B^0 \to K^{*\,0} \mu^+\mu^-$, $B_s \to \phi \mu^+\mu^-$, and $\Lambda_b \to \Lambda \mu^+\mu^-$ decays. For each decay mode, we take into account one broad $q^2$ bin below the narrow charmonium resonances and one broad bin above. Considering finer $q^2$ bins that are often also available has no advantage for the determination of the CKM matrix elements. 

The main theoretical uncertainties of the semileptonic branching ratio predictions arise from form factors and from additional non-factorizable effects. We use \verb|flavio| version~\verb|2.3.0|~\cite{Straub:2018kue} with default hadronic parameters to determine the central values and the uncertainties of the branching ratios, as well as the correlations of the uncertainties. 
The form factors implemented in \verb|flavio| are based on a combined fit of light-cone sum rule and lattice QCD results~\cite{Bharucha:2015bzk} (see also \cite{Gubernari:2018wyi}) and therefore lead to sizeable error correlations of the branching ratio predictions in the low-$q^2$ and high-$q^2$ bins. Moverover, there are non-negligible correlations between the uncertainties of the two pseudoscalar to vector transitions $B^0 \to K^{*\,0} \mu^+\mu^-$ and $B_s \to \phi \mu^+\mu^-$, due to the approximate $SU(3)$ flavor symmetry. 
We neglect additional percent level correlations between the uncertainties of the considered decay modes that are due to common input parameters.
For the numerical coefficients in~\eqref{eq:semileptonicBR}, we find for the $B^+ \to K^+ \mu^+\mu^-$ decay
\begin{equation}
    a_{B^+ \to K^+}^{[1.1,6]} = (1.00 \pm 0.16)\times 10^{-4} ~, \qquad a_{B^+ \to K^+}^{[15,22]} = (0.61 \pm 0.06)\times 10^{-4} ~,
\end{equation}
with an error correlation of $\rho = +68\%$. 
For the baryonic decay $\Lambda_b \to \Lambda \mu^+\mu^-$ we find 
\begin{equation}
    a_{\Lambda_b \to \Lambda}^{[1.1,6]} = (0.30 \pm 0.16)\times 10^{-4} ~, \qquad a_{\Lambda_b \to \Lambda}^{[15,20]} = (2.07 \pm 0.21)\times 10^{-4} ~,
\end{equation}
with a modest error correlation of $\rho = +5\%$. Finally, for the decays $B^0 \to K^{*\,0} \mu^+\mu^-$ and $B_s \to \phi \mu^+\mu^-$ we find the following coefficients and correlation matrix
\begin{equation}
    \begin{matrix} 
    a_{B^0 \to K^{*\,0}}^{[1.1,6]} =(1.36 \pm 0.16)\times 10^{-4} \\ 
    a_{B^0 \to K^{*\,0}}^{[15,19]} =(1.39 \pm 0.15)\times 10^{-4} \\ 
   ~~~ a_{B_s \to \phi}^{[1.1,6]} =(1.54 \pm 0.14)\times 10^{-4} \\
   ~~~ a_{B_s \to \phi}^{[15,19]} =(1.30 \pm 0.12)\times 10^{-4}
    \end{matrix}  ~, \qquad 
    \rho = \begin{pmatrix} 
    1 & 0.50 & 0.24 & 0.01 \\
    &1& 0.00 & 0.22   \\
    &&1& 0.36 \\
    &&& 1 
    \end{pmatrix} ~.
\end{equation}

On the experimental side, we use the latest results from the LHCb collaboration for the branching ratios of $B^+ \to K^+ \mu^+\mu^-$~\cite{Aaij:2014pli}, $B^0 \to K^{*\,0} \mu^+\mu^-$~\cite{Aaij:2016flj}, $B_s \to \phi \mu^+\mu^-$~\cite{LHCb:2021zwz}, and $\Lambda_b \to \Lambda \mu^+\mu^-$~\cite{LHCb:2015tgy}
\begin{eqnarray}
    && \text{BR}(B^+ \to K^+ \mu^+\mu^-)_\text{exp}^{[1.1,6]} = (1.19 \pm 0.07)\times 10^{-7}  ~, \\
    && \text{BR}(B^+ \to K^+ \mu^+\mu^-)_\text{exp}^{[15,22]} = (0.85 \pm 0.05)\times 10^{-7} ~, \\
    && \text{BR}(B^0 \to K^{*\,0} \mu^+\mu^-)_\text{exp}^{[1.1,6]} = (1.68\pm 0.15)\times 10^{-7} ~, \\
    && \text{BR}(B^0 \to K^{*\,0} \mu^+\mu^-)_\text{exp}^{[15,19]} = (1.74\pm 0.14)\times 10^{-7} ~, \\
    && \text{BR}(B_s \to \phi \mu^+\mu^-)_\text{exp}^{[1.1,6]} = (1.41\pm0.10)\times 10^{-7} ~, \\
    && \text{BR}(B_s \to \phi \mu^+\mu^-)_\text{exp}^{[15,19]} = (1.85\pm0.13)\times 10^{-7} ~, \\
    && \text{BR}(\Lambda_b \to \Lambda \mu^+\mu^-)_\text{exp}^{[1.1,6]} = (0.44\pm0.31)\times 10^{-7} ~, \\
    && \text{BR}(\Lambda_b \to \Lambda \mu^+\mu^-)_\text{exp}^{[15,20]} = (6.00\pm1.34)\times 10^{-7} ~.
\end{eqnarray}
We note that in most cases the experimental precision is better than $\sim 10\%$ and has already surpassed the precision of the SM predictions.

%%%%%%%%%%%%%%%%%%%%%%%%%%%%%%%%%%%%%%%%%%%%%%%%%%
\section{Results of the Fits} \label{sec:fits}
%%%%%%%%%%%%%%%%%%%%%%%%%%%%%%%%%%%%%%%%%%%%%%%%%%

Based on the above SM expressions for the various observables in terms of the CKM parameters $\lambda$, $|V_{cb}|$, $|V_{ub}|$, and $\gamma$, and the corresponding experimental measurements, we construct a $\chi^2$ function that takes into account the experimental uncertainties as well as the theoretical uncertainties and their correlations in the form of covariance matrices.

As the sine of the Cabibbo angle is known with excellent precision, we simply set it to its central value $\lambda \simeq 0.2248$ \cite{Charles:2004jd} in our numerical analysis. For the CKM angle $\gamma$, we use the HFLAV average $\gamma = (66.1 ^{+3.4}_{-3.6})^\circ$~\cite{Amhis:2019ckw}, which is dominated by the precise LHCb results from~\cite{LHCb:2020kho} (see also the very recent update~\cite{LHCb:2021dcr}). The importance of precise determinations of $\gamma$ has recently been emphasized also in~\cite{Blanke:2018cya, Buras:2021nns}. By profiling over $\gamma$ we arrive at best fit contours in the $|V_{cb}|$ - $|V_{ub}|$ plane.

%%%%%%%%%%%%%%%%%%%%%%%%%%%%%%%%%
\subsection{Loop level determination from meson oscillations} \label{sec:mixing_fit}
%%%%%%%%%%%%%%%%%%%%%%%%%%%%%%%%%

\begin{figure}[tb]
\centering
\includegraphics[width=0.6\textwidth]{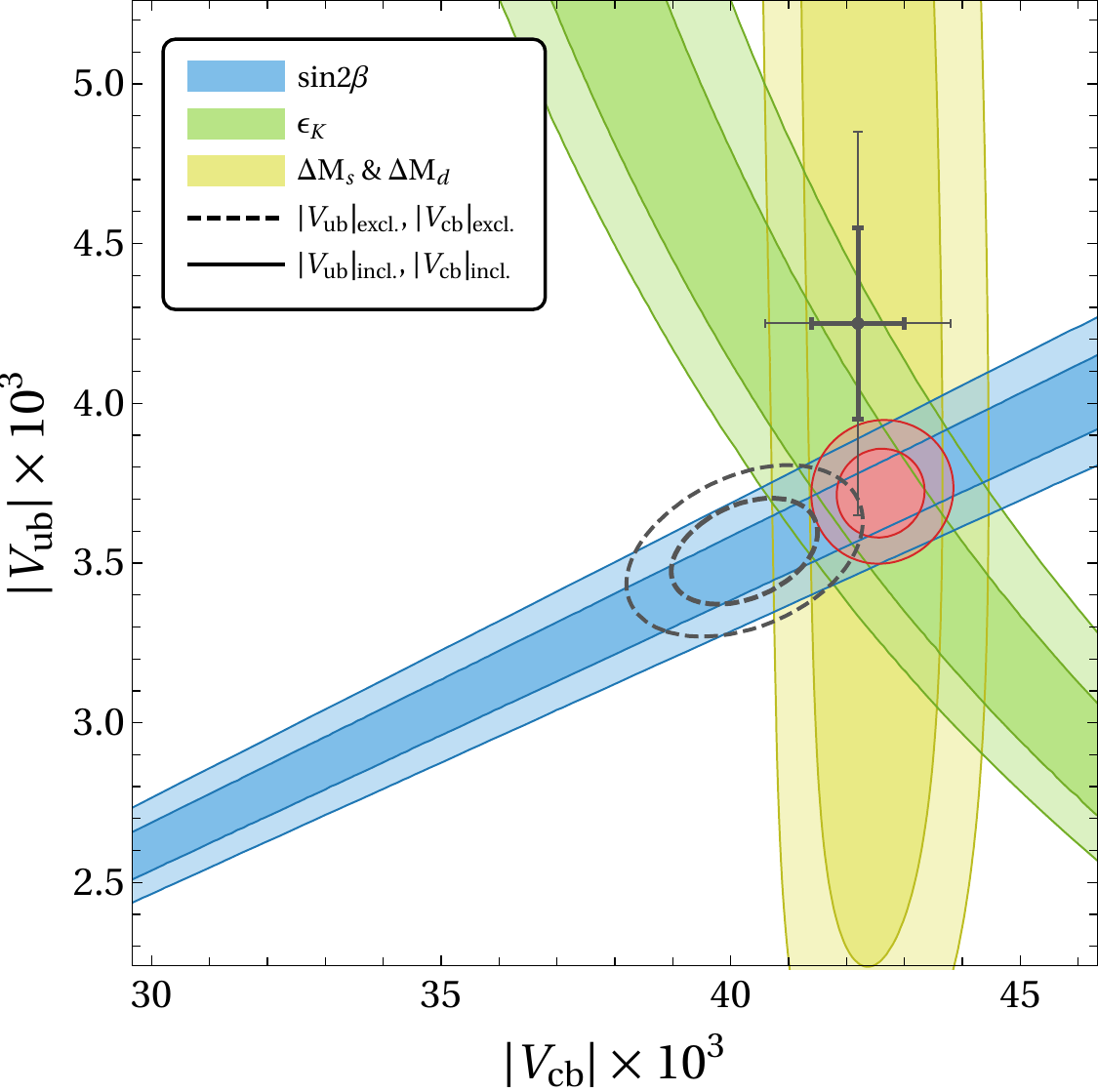} 
\caption{Constraints in the $|V_{cb}|$ - $|V_{ub}|$ plane from meson mixing observables. The blue, green, and yellow bands show the constraints from $\sin(2\beta)$, $\epsilon_K$, and $\Delta M_d$ and $\Delta M_s$, respectively. The combined best fit region is shown in red. The dashed contours and the gray error bars show for comparison the exclusive and inclusive values from Figure~\ref{fig:incl_excl}.}
\label{fig:mixing}
\end{figure}

We first focus on the meson oscillation observables, $\epsilon_K$, $\Delta M_d$, $\Delta M_s$, and $\sin(2\beta)$ as discussed in sections~\ref{sec:epsK} and~\ref{sec:Bmixing}. In Figure~\ref{fig:mixing} we show the individual $1\sigma$ and $2\sigma$ constraints in the $|V_{cb}|$ - $|V_{ub}|$ plane from $\epsilon_K$ (green), from $\sin(2\beta)$ (blue), and from the combination of $\Delta M_d$ and $\Delta M_s$ (yellow). The red region is the combination of those constraints and corresponds to the values 
\begin{equation}
    |V_{cb}|_\text{meson mixing} = (42.6 \pm 0.5)\times 10^{-3}~, \qquad |V_{ub}|_\text{meson mixing} = (3.72 \pm 0.09) \times 10^{-3}~.
\end{equation}
with a negligibly small error correlation. These central values are very close to the results from global CKM fits~\cite{Charles:2004jd,UTfit:2006vpt}. Our values of $|V_{cb}|$ and $|V_{ub}|$ from meson mixing observables also agree very well with the ones recently found in~\cite{Buras:2021nns}.

%%%%%%%%%%%%%%%%%%%%%%%%%%%%%%%%%%%%%%%%
%%%%%%%%%%%%%%%%%%%%%%%%%%%%%%%%%%%%%%%%

We observe that the $B$ meson oscillation frequencies $\Delta M_d$ and $\Delta M_s$ predict a value for $|V_{cb}|$ that is in excellent agreement with the inclusive determination.\footnote{As mentioned in section~\ref{sec:Bmixing}, we use the $N_f = 2+1+1$ lattice results for the hadronic matrix elements relevant for $B_s$ and $B^0$ mixing~\cite{Aoki:2021kgd}. Comparable results are obtained if sum rule determinations of the hadronic matrix elements are used~\cite{King:2019rvk}. We checked that the preferred value for $|V_{cb}|$ is very close to the exclusive determination if we use the $N_f = 2+1$ lattice results.} Also the combination of the CP violating observables $\epsilon_K$ and $\sin(2\beta)$ prefers a value for $|V_{cb}|$ close to the inclusive determination. The combination of all meson mixing observables shows a remarkable level of consistency. The combination is dominated by $\epsilon_K$ and $\sin(2\beta)$, while $\Delta M_d$ and $\Delta M_s$ play a slightly lesser role due to the slightly larger theory uncertainties. Continued improvements in the determination of the hadronic $B$ mixing matrix elements are thus highly motivated. 

Overall, we observe that our combination of meson mixing observables is compatible with both the inclusive and the exclusive determination of $|V_{cb}|$ and $|V_{ub}|$ to better than $2\sigma$. There is a slight preference for the inclusive value of $|V_{cb}|$ and the exclusive value of $|V_{ub}|$.

%%%%%%%%%%%%%%%%%%%%%%%%%%%%%%%%%
\subsection{Loop level determination from rare b decay rates} \label{sec:rare_fit}
%%%%%%%%%%%%%%%%%%%%%%%%%%%%%%%%%

The rare $B$ decays discussed in section~\ref{sec:rareB} can be used to determine $|V_{cb}|$ but have only very weak sensitivity to the value of $|V_{ub}|$. We therefore do not show the results in the $|V_{cb}|$ - $|V_{ub}|$ plane but directly as values for $|V_{cb}|$, profiling over $|V_{ub}|$ within the exclusive value from the PDG, cf. equation~\eqref{eq:VcbVub_excl}. We checked that using instead the inclusive value for $|V_{ub}|$ leads to negligible differences in our results. 

\begin{figure}[tb]
\centering
\includegraphics[width=0.75\textwidth]{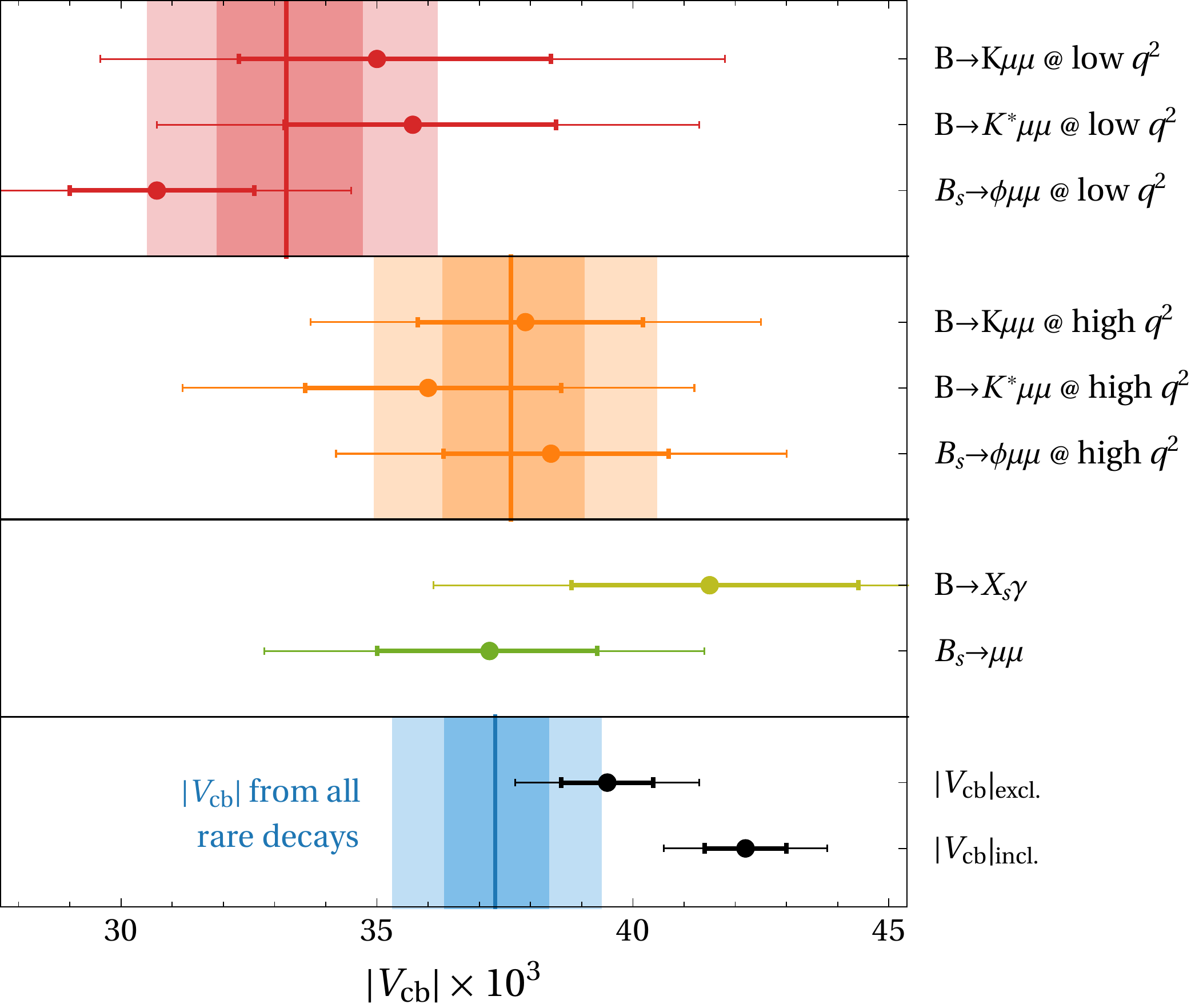} 
\caption{Determinations of $|V_{cb}|$ from various rare decays of $B$ mesons. The error bars correspond to the individual $1\sigma$ and $2\sigma$ uncertainties. The red and orange bands show the combination of the semileptonic decays at low $q^2$ and high $q^2$, respectively. The blue band is the combination of all rare $B$ decay data. For comparison, the PDG averages of the inclusive and exclusive determinations of $|V_{cb}|$ are shown as well.}
\label{fig:raredecays}
\end{figure}

The error bars in Figure~\ref{fig:raredecays} show the best fit values for $|V_{cb}|$ based on the rare semileptonic decays $B^+ \to K^+ \mu^+ \mu^-$, $B^0 \to K^{*\,0} \mu^+ \mu^-$ and $B_s \to \phi \mu^+\mu^-$ at low $q^2$ (red) and at high $q^2$ (orange), as well as from $B \to X_s \gamma$ (yellow) and $B_s \to \mu^+ \mu^-$ (green). The colored bands in the figure show the results from the combinations of the semileptonic decays at low $q^2$ and high $q^2$ as well as the combination of all rare $B$ decay results. The results from the baryonic decay $\Lambda_b \to \Lambda \mu^+\mu^-$ are not shown individually as they have large uncertainties but they are included in the combinations. The exclusive and inclusive values for $|V_{cb}|$ from the PDG are shown for comparison. 

%%%%%%%%%%%%%%%%%%%%%%%%%%%%%%%%%%%%%%%%
%%%%%%%%%%%%%%%%%%%%%%%%%%%%%%%%%%%%%%%%

As is well known, the experimental results for the branching ratios of $B \to K \mu^+ \mu^-$, $B \to K^{*} \mu^+ \mu^-$ and $B_s \to \phi \mu^+\mu^-$~\cite{Aaij:2014pli, Aaij:2016flj, LHCb:2021zwz} are all significantly low compared to SM predictions, in particular in the low $q^2$ region. These results have been interpreted as possible signs of new physics effects in rare $B$ decays. In the context of the SM, the results instead are an indication for a very small value of $|V_{cb}|$. For example, the $B_s \to \phi \mu^+\mu^-$ branching ratio at low $q^2$ points to a particularly low central value $|V_{cb}| \simeq 31 \times 10^{-3}$, much lower than the inclusive or exclusive tree level determinations.
The results of our combinations of the semileptonic decays at low $q^2$ and high $q^2$ are
\begin{equation}
    |V_{cb}|_{\text{low}~q^2} = (33.2 \pm 1.5)\times 10^{-3}~, \qquad |V_{cb}|_{\text{high}~q^2} = (37.6 \pm 1.4)\times 10^{-3}~.
\end{equation}
While the result from high $q^2$ is compatible with the tree level exclusive determination, it is more than $2\sigma$ below the value from the inclusive determination. Our result from low $q^2$ is more than $3\sigma$ below the exclusive value and more than $5\sigma$ below the inclusive one. A small $|V_{cb}|$ thus appears to be an unlikely explanation of the low $b \to s \mu^+ \mu^-$ branching ratio data.

For comparison, the value for $|V_{cb}|$ we find from the $B \to X_s \gamma$ decay sits between the inclusive and exclusive determinations and has a large uncertainty. It is compatible with both determinations to better than $1\sigma$. 
The $B_s \to \mu^+ \mu^-$ decay gives a $|V_{cb}|$ that is approximately $2\sigma$ below the inclusive determination. Interestingly, it is compatible with the exclusive determination to better than $1\sigma$. This indicates that the $2\sigma$ tension that is observed between the SM prediction of BR$(B_s \to \mu^+ \mu^-)$~\cite{Beneke:2019slt} and the experimental measurements~\cite{LHCb:2021awg, LHCb:2021vsc, Aaboud:2018mst, Sirunyan:2019xdu} can be largely resolved if the exclusive determination of $|V_{cb}|$ were used for the SM prediction. This has been also pointed out in~\cite{Bobeth:2021cxm}. 

Combining all rare $B$ decay data, we arrive at the average
\begin{equation}
    |V_{cb}|_\text{rare decays} = (37.3 \pm 1.0)\times 10^{-3}~.
\end{equation}
The central value is significantly below the inclusive determination of $|V_{cb}|$. It is also below the exclusive determination, but compatible to better than $2 \sigma$. Interestingly, the uncertainty is only slightly larger than the uncertainty of the tree level determination. However, the uncertainty is dominated by theory and challenging to improve.

%%%%%%%%%%%%%%%%%%%%%%%%%%%%%%%%%
\subsection{Global fit of $|V_{cb}|$ and $|V_{ub}|$ from loops}
%%%%%%%%%%%%%%%%%%%%%%%%%%%%%%%%%

\begin{figure}[tb]
\centering
\includegraphics[width=0.6\textwidth]{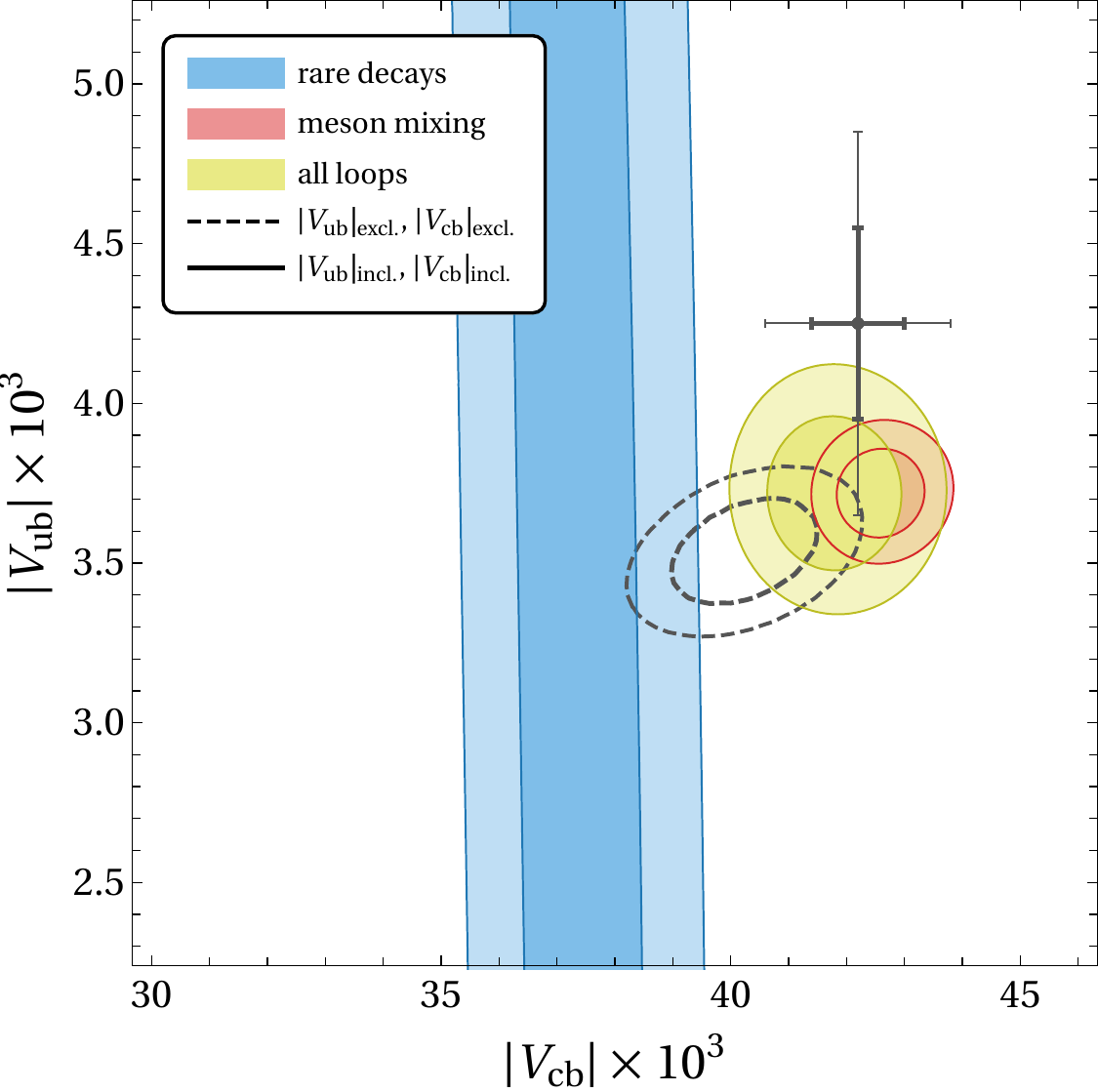} 
\caption{Comparison of all $|V_{cb}|$ and $|V_{ub}|$ determinations discussed in this work. The black cross shows the inclusive values from the PDG, cf. equation \eqref{eq:VcbVub_incl}. The dashed ellipse is our combination of all exclusive tree level determinations as discussed in section~\ref{sec:direct}. The red region corresponds to our fit of meson oscillation data from section~\ref{sec:mixing_fit}. The blue band corresponds to the determination using rare $B$ decays from section~\ref{sec:rare_fit}. Finally, the yellow region corresponds to a combined fit of all loop determinations with error contours inflated to account for the tensions in the fit.}
\label{fig:global}
\end{figure}

Finally, we combine the various loop determinations of $|V_{cb}|$ and $|V_{ub}|$ in a global fit. We take into account the meson mixing observables discussed in sections~\ref{sec:epsK} and~\ref{sec:Bmixing} and the rare meson decays discussed in section~\ref{sec:rareB} and compare to the tree level determinations from section~\ref{sec:direct}. The result is shown in Figure~\ref{fig:global}. 

%%%%%%%%%%%%%%%%%%%%%%%%%%%%%%%%%%%%%%%%
%%%%%%%%%%%%%%%%%%%%%%%%%%%%%%%%%%%%%%%%

We observe considerable tensions between the different determinations. In particular, there is a large tension between the rare decays (blue band) on the one side and the meson mixing observables (red ellipse) and the tree level inclusive determination (black error bars) on the other side. Our average of the tree level exclusive determinations (dashed ellipse) sits in between. To obtain the global loop combination shown in yellow, we follow the PDG prescription and inflate the uncertainty by $\sqrt{\chi^2_\text{bf} / N_\text{dof}} = \sqrt{40.5/13} = 1.76$. Employing a Gaussian approximation, we find
\begin{equation}
    |V_{cb}|_\text{loop} = (41.75 \pm 0.76)\times 10^{-3}~, \qquad |V_{ub}|_\text{loop} = (3.71 \pm 0.16) \times 10^{-3}~.
    \end{equation}
with a negligible error correlation.
The central values are in very good agreement with the results from global CKM fits~\cite{Charles:2004jd,UTfit:2006vpt} that do not take into account rare $B$ decay data.
The uncertainties of our loop level determinations are considerably larger, due to the omission of the tree level information and the error inflation mentioned above.

%%%%%%%%%%%%%%%%%%%%%%%%%%%%%%%%%
\section{Conclusions} \label{sec:conclusions}
%%%%%%%%%%%%%%%%%%%%%%%%%%%%%%%%%

Precise determinations of $V_{cb}$ and $V_{ub}$ are crucial inputs for testing the SM CKM picture of flavor and CP violation. 
For many years, discrepancies between determinations using exclusive and inclusive tree level $B$ decays have limited the precision of $V_{cb}$ and $V_{ub}$ determinations. A summary of the current status is provided in Figure~\ref{fig:incl_excl}.
In this paper, we used loop level processes to obtain the absolute values of $V_{cb}$ and $V_{ub}$. Such a loop level strategy gives valid results in the absence of new physics effects in the considered observables. We focused on two classes of loop processes: neutral meson oscillations and rare $b$ hadron decays. 

Our main results are summarized in Figures~\ref{fig:mixing} and~\ref{fig:raredecays}. We find that the combination of the observables $\epsilon_K$ (that parameterizes indirect CP violation in Kaon oscillations) and $\sin(2\beta)$ (that parameterizes indirect CP violation in $B^0$ oscillations) gives a precise determination of $|V_{cb}|$ and $|V_{ub}|$, with central values close to the inclusive value of $|V_{cb}|$ and the exclusive value for $|V_{ub}|$. Also the $B$ meson oscillation frequencies $\Delta M_d$ and $\Delta M_s$ show preference for the inclusive value of $|V_{cb}|$, but are also compatible with the exlusive value, given the current theory errors. To improve the constraining power of $\Delta M_d$ and $\Delta M_s$, more precise determinations of the hadronic matrix elements in $B$ meson mixing are required. 

As is well known, the experimental measurements of several rare $B$ decay branching ratios are significantly below the SM predictions. Therefore, $|V_{cb}|$ determinations based on rare $b$ decay data give values far below the direct determinations from tree level decays. The preference for small $|V_{cb}|$ is particularly pronounced if one focuses on rare semileptonic $B$ decays at low di-lepton invariant mass. In that case we find a value for $|V_{cb}|$ that is more than $3\sigma$ below the exclusive value and more than $5\sigma$ below the inclusive one. We thus consider it unlikely that $|V_{cb}|$ is the sole culprit behind the low $b \to s \mu^+ \mu^-$ branching ratios.

Interestingly, the $\sim 2\sigma$ tension between the commonly quoted SM prediction for the $B_s \to \mu^+ \mu^-$ branching ratio and the corresponding experimental world average largely disappears if the exclusive value of $|V_{cb}|$ were used in the SM prediction.

%%%%%%%%%%%%%%%%%%%%%%%%%%%%%%%%%
\section*{Acknowledgements}
%%%%%%%%%%%%%%%%%%%%%%%%%%%%%%%%%

We thank Andrzej Buras and Alexander Lenz for useful feedback.
The research of WA was supported by the U.S. Department of Energy grant number DE-SC0010107.

%%%%%%%%%%%%%%%%%%%%%%%%%%%%%%%%%
\bibliography{VubVcb}

\begin{thebibliography}{71}%
\makeatletter
\providecommand \@ifxundefined [1]{%
 \@ifx{#1\undefined}
}%
\providecommand \@ifnum [1]{%
 \ifnum #1\expandafter \@firstoftwo
 \else \expandafter \@secondoftwo
 \fi
}%
\providecommand \@ifx [1]{%
 \ifx #1\expandafter \@firstoftwo
 \else \expandafter \@secondoftwo
 \fi
}%
\providecommand \natexlab [1]{#1}%
\providecommand \enquote  [1]{``#1''}%
\providecommand \bibnamefont  [1]{#1}%
\providecommand \bibfnamefont [1]{#1}%
\providecommand \citenamefont [1]{#1}%
\providecommand \href@noop [0]{\@secondoftwo}%
\providecommand \href [0]{\begingroup \@sanitize@url \@href}%
\providecommand \@href[1]{\@@startlink{#1}\@@href}%
\providecommand \@@href[1]{\endgroup#1\@@endlink}%
\providecommand \@sanitize@url [0]{\catcode `\\12\catcode `\$12\catcode
  `\&12\catcode `\#12\catcode `\^12\catcode `\_12\catcode `\%12\relax}%
\providecommand \@@startlink[1]{}%
\providecommand \@@endlink[0]{}%
\providecommand \url  [0]{\begingroup\@sanitize@url \@url }%
\providecommand \@url [1]{\endgroup\@href {#1}{\urlprefix }}%
\providecommand \urlprefix  [0]{URL }%
\providecommand \Eprint [0]{\href }%
\providecommand \doibase [0]{https://doi.org/}%
\providecommand \selectlanguage [0]{\@gobble}%
\providecommand \bibinfo  [0]{\@secondoftwo}%
\providecommand \bibfield  [0]{\@secondoftwo}%
\providecommand \translation [1]{[#1]}%
\providecommand \BibitemOpen [0]{}%
\providecommand \bibitemStop [0]{}%
\providecommand \bibitemNoStop [0]{.\EOS\space}%
\providecommand \EOS [0]{\spacefactor3000\relax}%
\providecommand \BibitemShut  [1]{\csname bibitem#1\endcsname}%
\let\auto@bib@innerbib\@empty
%</preamble>
\bibitem [{\citenamefont {Cabibbo}(1963)}]{Cabibbo:1963yz}%
  \BibitemOpen
  \bibfield  {author} {\bibinfo {author} {\bibfnamefont {N.}~\bibnamefont
  {Cabibbo}},\ }\bibfield  {title} {\bibinfo {title} {{Unitary Symmetry and
  Leptonic Decays}},\ }\href {https://doi.org/10.1103/PhysRevLett.10.531}
  {\bibfield  {journal} {\bibinfo  {journal} {Phys. Rev. Lett.}\ }\textbf
  {\bibinfo {volume} {10}},\ \bibinfo {pages} {531} (\bibinfo {year}
  {1963})}\BibitemShut {NoStop}%
\bibitem [{\citenamefont {Kobayashi}\ and\ \citenamefont
  {Maskawa}(1973)}]{Kobayashi:1973fv}%
  \BibitemOpen
  \bibfield  {author} {\bibinfo {author} {\bibfnamefont {M.}~\bibnamefont
  {Kobayashi}}\ and\ \bibinfo {author} {\bibfnamefont {T.}~\bibnamefont
  {Maskawa}},\ }\bibfield  {title} {\bibinfo {title} {{CP Violation in the
  Renormalizable Theory of Weak Interaction}},\ }\href
  {https://doi.org/10.1143/PTP.49.652} {\bibfield  {journal} {\bibinfo
  {journal} {Prog. Theor. Phys.}\ }\textbf {\bibinfo {volume} {49}},\ \bibinfo
  {pages} {652} (\bibinfo {year} {1973})}\BibitemShut {NoStop}%
\bibitem [{\citenamefont {Ricciardi}\ and\ \citenamefont
  {Rotondo}(2020)}]{Ricciardi:2019zph}%
  \BibitemOpen
  \bibfield  {author} {\bibinfo {author} {\bibfnamefont {G.}~\bibnamefont
  {Ricciardi}}\ and\ \bibinfo {author} {\bibfnamefont {M.}~\bibnamefont
  {Rotondo}},\ }\bibfield  {title} {\bibinfo {title} {{Determination of the
  Cabibbo-Kobayashi-Maskawa matrix element $|V_{cb}|$}},\ }\href
  {https://doi.org/10.1088/1361-6471/ab9f01} {\bibfield  {journal} {\bibinfo
  {journal} {J. Phys. G}\ }\textbf {\bibinfo {volume} {47}},\ \bibinfo {pages}
  {113001} (\bibinfo {year} {2020})},\ \Eprint
  {https://arxiv.org/abs/1912.09562} {arXiv:1912.09562 [hep-ph]} \BibitemShut
  {NoStop}%
\bibitem [{\citenamefont {Ricciardi}(2021)}]{Ricciardi:2021shl}%
  \BibitemOpen
  \bibfield  {author} {\bibinfo {author} {\bibfnamefont {G.}~\bibnamefont
  {Ricciardi}},\ }\bibfield  {title} {\bibinfo {title} {{Semileptonic $B$
  decays and $|V_{xb}|$ update}},\ }in\ \href@noop {} {\emph {\bibinfo
  {booktitle} {{19th International Conference on B-Physics at Frontier
  Machines}}}}\ (\bibinfo {year} {2021})\ \Eprint
  {https://arxiv.org/abs/2103.06099} {arXiv:2103.06099 [hep-ph]} \BibitemShut
  {NoStop}%
\bibitem [{\citenamefont {Zyla}\ \emph {et~al.}(2020)\citenamefont {Zyla} \emph
  {et~al.}}]{Zyla:2020zbs}%
  \BibitemOpen
  \bibfield  {author} {\bibinfo {author} {\bibfnamefont {P.~A.}\ \bibnamefont
  {Zyla}} \emph {et~al.} (\bibinfo {collaboration} {Particle Data Group}),\
  }\bibfield  {title} {\bibinfo {title} {{Review of Particle Physics}},\ }\href
  {https://doi.org/10.1093/ptep/ptaa104} {\bibfield  {journal} {\bibinfo
  {journal} {PTEP}\ }\textbf {\bibinfo {volume} {2020}},\ \bibinfo {pages}
  {083C01} (\bibinfo {year} {2020})}\BibitemShut {NoStop}%
\bibitem [{\citenamefont {Kang}\ \emph {et~al.}(2014)\citenamefont {Kang},
  \citenamefont {Kubis}, \citenamefont {Hanhart},\ and\ \citenamefont
  {Mei\ss{}ner}}]{Kang:2013jaa}%
  \BibitemOpen
  \bibfield  {author} {\bibinfo {author} {\bibfnamefont {X.-W.}\ \bibnamefont
  {Kang}}, \bibinfo {author} {\bibfnamefont {B.}~\bibnamefont {Kubis}},
  \bibinfo {author} {\bibfnamefont {C.}~\bibnamefont {Hanhart}},\ and\ \bibinfo
  {author} {\bibfnamefont {U.-G.}\ \bibnamefont {Mei\ss{}ner}},\ }\bibfield
  {title} {\bibinfo {title} {{$B_{l4}$ decays and the extraction of
  $|V_{ub}|$}},\ }\href {https://doi.org/10.1103/PhysRevD.89.053015} {\bibfield
   {journal} {\bibinfo  {journal} {Phys. Rev. D}\ }\textbf {\bibinfo {volume}
  {89}},\ \bibinfo {pages} {053015} (\bibinfo {year} {2014})},\ \Eprint
  {https://arxiv.org/abs/1312.1193} {arXiv:1312.1193 [hep-ph]} \BibitemShut
  {NoStop}%
\bibitem [{\citenamefont {Crivellin}\ and\ \citenamefont
  {Pokorski}(2015)}]{Crivellin:2014zpa}%
  \BibitemOpen
  \bibfield  {author} {\bibinfo {author} {\bibfnamefont {A.}~\bibnamefont
  {Crivellin}}\ and\ \bibinfo {author} {\bibfnamefont {S.}~\bibnamefont
  {Pokorski}},\ }\bibfield  {title} {\bibinfo {title} {{Can the differences in
  the determinations of $V_{ub}$ and $V_{cb}$ be explained by New Physics?}},\
  }\href {https://doi.org/10.1103/PhysRevLett.114.011802} {\bibfield  {journal}
  {\bibinfo  {journal} {Phys. Rev. Lett.}\ }\textbf {\bibinfo {volume} {114}},\
  \bibinfo {pages} {011802} (\bibinfo {year} {2015})},\ \Eprint
  {https://arxiv.org/abs/1407.1320} {arXiv:1407.1320 [hep-ph]} \BibitemShut
  {NoStop}%
\bibitem [{\citenamefont {Bernlochner}\ \emph {et~al.}(2014)\citenamefont
  {Bernlochner}, \citenamefont {Ligeti},\ and\ \citenamefont
  {Turczyk}}]{Bernlochner:2014ova}%
  \BibitemOpen
  \bibfield  {author} {\bibinfo {author} {\bibfnamefont {F.~U.}\ \bibnamefont
  {Bernlochner}}, \bibinfo {author} {\bibfnamefont {Z.}~\bibnamefont
  {Ligeti}},\ and\ \bibinfo {author} {\bibfnamefont {S.}~\bibnamefont
  {Turczyk}},\ }\bibfield  {title} {\bibinfo {title} {{New ways to search for
  right-handed current in $B \to \rho \ell \nu$ decay}},\ }\href
  {https://doi.org/10.1103/PhysRevD.90.094003} {\bibfield  {journal} {\bibinfo
  {journal} {Phys. Rev. D}\ }\textbf {\bibinfo {volume} {90}},\ \bibinfo
  {pages} {094003} (\bibinfo {year} {2014})},\ \Eprint
  {https://arxiv.org/abs/1408.2516} {arXiv:1408.2516 [hep-ph]} \BibitemShut
  {NoStop}%
\bibitem [{\citenamefont {Colangelo}\ and\ \citenamefont
  {De~Fazio}(2017)}]{Colangelo:2016ymy}%
  \BibitemOpen
  \bibfield  {author} {\bibinfo {author} {\bibfnamefont {P.}~\bibnamefont
  {Colangelo}}\ and\ \bibinfo {author} {\bibfnamefont {F.}~\bibnamefont
  {De~Fazio}},\ }\bibfield  {title} {\bibinfo {title} {{Tension in the
  inclusive versus exclusive determinations of $|V_{cb}|$: a possible role of
  new physics}},\ }\href {https://doi.org/10.1103/PhysRevD.95.011701}
  {\bibfield  {journal} {\bibinfo  {journal} {Phys. Rev. D}\ }\textbf {\bibinfo
  {volume} {95}},\ \bibinfo {pages} {011701} (\bibinfo {year} {2017})},\
  \Eprint {https://arxiv.org/abs/1611.07387} {arXiv:1611.07387 [hep-ph]}
  \BibitemShut {NoStop}%
\bibitem [{\citenamefont {Gambino}\ and\ \citenamefont
  {Hashimoto}(2020)}]{Gambino:2020crt}%
  \BibitemOpen
  \bibfield  {author} {\bibinfo {author} {\bibfnamefont {P.}~\bibnamefont
  {Gambino}}\ and\ \bibinfo {author} {\bibfnamefont {S.}~\bibnamefont
  {Hashimoto}},\ }\bibfield  {title} {\bibinfo {title} {{Inclusive Semileptonic
  Decays from Lattice QCD}},\ }\href
  {https://doi.org/10.1103/PhysRevLett.125.032001} {\bibfield  {journal}
  {\bibinfo  {journal} {Phys. Rev. Lett.}\ }\textbf {\bibinfo {volume} {125}},\
  \bibinfo {pages} {032001} (\bibinfo {year} {2020})},\ \Eprint
  {https://arxiv.org/abs/2005.13730} {arXiv:2005.13730 [hep-lat]} \BibitemShut
  {NoStop}%
\bibitem [{\citenamefont {Ferlewicz}\ \emph {et~al.}(2021)\citenamefont
  {Ferlewicz}, \citenamefont {Urquijo},\ and\ \citenamefont
  {Waheed}}]{Ferlewicz:2020lxm}%
  \BibitemOpen
  \bibfield  {author} {\bibinfo {author} {\bibfnamefont {D.}~\bibnamefont
  {Ferlewicz}}, \bibinfo {author} {\bibfnamefont {P.}~\bibnamefont {Urquijo}},\
  and\ \bibinfo {author} {\bibfnamefont {E.}~\bibnamefont {Waheed}},\
  }\bibfield  {title} {\bibinfo {title} {{Revisiting fits to $B^{0} \to D^{*-}
  \ell^{+} \nu_{\ell}$ to measure $|V_{cb}|$ with novel methods and preliminary
  LQCD data at nonzero recoil}},\ }\href
  {https://doi.org/10.1103/PhysRevD.103.073005} {\bibfield  {journal} {\bibinfo
   {journal} {Phys. Rev. D}\ }\textbf {\bibinfo {volume} {103}},\ \bibinfo
  {pages} {073005} (\bibinfo {year} {2021})},\ \Eprint
  {https://arxiv.org/abs/2008.09341} {arXiv:2008.09341 [hep-ph]} \BibitemShut
  {NoStop}%
\bibitem [{\citenamefont {Fael}\ \emph
  {et~al.}(2021{\natexlab{a}})\citenamefont {Fael}, \citenamefont
  {Sch\"onwald},\ and\ \citenamefont {Steinhauser}}]{Fael:2020njb}%
  \BibitemOpen
  \bibfield  {author} {\bibinfo {author} {\bibfnamefont {M.}~\bibnamefont
  {Fael}}, \bibinfo {author} {\bibfnamefont {K.}~\bibnamefont {Sch\"onwald}},\
  and\ \bibinfo {author} {\bibfnamefont {M.}~\bibnamefont {Steinhauser}},\
  }\bibfield  {title} {\bibinfo {title} {{Relation between the
  $\overline{\mathrm{MS}}$ and the kinetic mass of heavy quarks}},\ }\href
  {https://doi.org/10.1103/PhysRevD.103.014005} {\bibfield  {journal} {\bibinfo
   {journal} {Phys. Rev. D}\ }\textbf {\bibinfo {volume} {103}},\ \bibinfo
  {pages} {014005} (\bibinfo {year} {2021}{\natexlab{a}})},\ \Eprint
  {https://arxiv.org/abs/2011.11655} {arXiv:2011.11655 [hep-ph]} \BibitemShut
  {NoStop}%
\bibitem [{\citenamefont {Fael}\ \emph
  {et~al.}(2021{\natexlab{b}})\citenamefont {Fael}, \citenamefont
  {Sch\"onwald},\ and\ \citenamefont {Steinhauser}}]{Fael:2020tow}%
  \BibitemOpen
  \bibfield  {author} {\bibinfo {author} {\bibfnamefont {M.}~\bibnamefont
  {Fael}}, \bibinfo {author} {\bibfnamefont {K.}~\bibnamefont {Sch\"onwald}},\
  and\ \bibinfo {author} {\bibfnamefont {M.}~\bibnamefont {Steinhauser}},\
  }\bibfield  {title} {\bibinfo {title} {{Third order corrections to the
  semileptonic $b\to c$ and the muon decays}},\ }\href
  {https://doi.org/10.1103/PhysRevD.104.016003} {\bibfield  {journal} {\bibinfo
   {journal} {Phys. Rev. D}\ }\textbf {\bibinfo {volume} {104}},\ \bibinfo
  {pages} {016003} (\bibinfo {year} {2021}{\natexlab{b}})},\ \Eprint
  {https://arxiv.org/abs/2011.13654} {arXiv:2011.13654 [hep-ph]} \BibitemShut
  {NoStop}%
\bibitem [{\citenamefont {Capdevila}\ \emph {et~al.}(2021)\citenamefont
  {Capdevila}, \citenamefont {Gambino},\ and\ \citenamefont
  {Nandi}}]{Capdevila:2021vkf}%
  \BibitemOpen
  \bibfield  {author} {\bibinfo {author} {\bibfnamefont {B.}~\bibnamefont
  {Capdevila}}, \bibinfo {author} {\bibfnamefont {P.}~\bibnamefont {Gambino}},\
  and\ \bibinfo {author} {\bibfnamefont {S.}~\bibnamefont {Nandi}},\ }\bibfield
   {title} {\bibinfo {title} {{Perturbative corrections to power suppressed
  effects in $\bar B\to X_u\ell\nu$}},\ }\href
  {https://doi.org/10.1007/JHEP04(2021)137} {\bibfield  {journal} {\bibinfo
  {journal} {JHEP}\ }\textbf {\bibinfo {volume} {04}},\ \bibinfo {pages}
  {137}},\ \Eprint {https://arxiv.org/abs/2102.03343} {arXiv:2102.03343
  [hep-ph]} \BibitemShut {NoStop}%
\bibitem [{\citenamefont {Leljak}\ \emph {et~al.}(2021)\citenamefont {Leljak},
  \citenamefont {Meli\'c},\ and\ \citenamefont {van Dyk}}]{Leljak:2021vte}%
  \BibitemOpen
  \bibfield  {author} {\bibinfo {author} {\bibfnamefont {D.}~\bibnamefont
  {Leljak}}, \bibinfo {author} {\bibfnamefont {B.}~\bibnamefont {Meli\'c}},\
  and\ \bibinfo {author} {\bibfnamefont {D.}~\bibnamefont {van Dyk}},\
  }\bibfield  {title} {\bibinfo {title} {{The $ \bar B \to \pi$ form factors
  from QCD and their impact on $|V_{ub}|$}},\ }\href
  {https://doi.org/10.1007/JHEP07(2021)036} {\bibfield  {journal} {\bibinfo
  {journal} {JHEP}\ }\textbf {\bibinfo {volume} {07}},\ \bibinfo {pages}
  {036}},\ \Eprint {https://arxiv.org/abs/2102.07233} {arXiv:2102.07233
  [hep-ph]} \BibitemShut {NoStop}%
\bibitem [{\citenamefont {Biswas}\ \emph {et~al.}(2021)\citenamefont {Biswas},
  \citenamefont {Nandi}, \citenamefont {Patra},\ and\ \citenamefont
  {Ray}}]{Biswas:2021qyq}%
  \BibitemOpen
  \bibfield  {author} {\bibinfo {author} {\bibfnamefont {A.}~\bibnamefont
  {Biswas}}, \bibinfo {author} {\bibfnamefont {S.}~\bibnamefont {Nandi}},
  \bibinfo {author} {\bibfnamefont {S.~K.}\ \bibnamefont {Patra}},\ and\
  \bibinfo {author} {\bibfnamefont {I.}~\bibnamefont {Ray}},\ }\bibfield
  {title} {\bibinfo {title} {{A closer look at the extraction of $|V_{ub}|$
  from $B \to \pi \ell \nu$}},\ }\href
  {https://doi.org/10.1007/JHEP07(2021)082} {\bibfield  {journal} {\bibinfo
  {journal} {JHEP}\ }\textbf {\bibinfo {volume} {07}},\ \bibinfo {pages}
  {082}},\ \Eprint {https://arxiv.org/abs/2103.01809} {arXiv:2103.01809
  [hep-ph]} \BibitemShut {NoStop}%
\bibitem [{\citenamefont {Mannel}\ \emph {et~al.}(2021)\citenamefont {Mannel},
  \citenamefont {Rahimi},\ and\ \citenamefont {Vos}}]{Mannel:2021mwe}%
  \BibitemOpen
  \bibfield  {author} {\bibinfo {author} {\bibfnamefont {T.}~\bibnamefont
  {Mannel}}, \bibinfo {author} {\bibfnamefont {M.}~\bibnamefont {Rahimi}},\
  and\ \bibinfo {author} {\bibfnamefont {K.~K.}\ \bibnamefont {Vos}},\
  }\bibfield  {title} {\bibinfo {title} {{Impact of background effects on the
  inclusive V$_{cb}$ determination}},\ }\href
  {https://doi.org/10.1007/JHEP09(2021)051} {\bibfield  {journal} {\bibinfo
  {journal} {JHEP}\ }\textbf {\bibinfo {volume} {09}},\ \bibinfo {pages}
  {051}},\ \Eprint {https://arxiv.org/abs/2105.02163} {arXiv:2105.02163
  [hep-ph]} \BibitemShut {NoStop}%
\bibitem [{\citenamefont {Martinelli}\ \emph {et~al.}(2021)\citenamefont
  {Martinelli}, \citenamefont {Simula},\ and\ \citenamefont
  {Vittorio}}]{Martinelli:2021onb}%
  \BibitemOpen
  \bibfield  {author} {\bibinfo {author} {\bibfnamefont {G.}~\bibnamefont
  {Martinelli}}, \bibinfo {author} {\bibfnamefont {S.}~\bibnamefont {Simula}},\
  and\ \bibinfo {author} {\bibfnamefont {L.}~\bibnamefont {Vittorio}},\
  }\bibfield  {title} {\bibinfo {title} {{$\vert V_{cb} \vert$ and $R(D^{(*)})$
  using lattice QCD and unitarity}},\ }\href@noop {} {\  (\bibinfo {year}
  {2021})},\ \Eprint {https://arxiv.org/abs/2105.08674} {arXiv:2105.08674
  [hep-ph]} \BibitemShut {NoStop}%
\bibitem [{\citenamefont {Bazavov}\ \emph {et~al.}(2021)\citenamefont {Bazavov}
  \emph {et~al.}}]{FermilabLattice:2021cdg}%
  \BibitemOpen
  \bibfield  {author} {\bibinfo {author} {\bibfnamefont {A.}~\bibnamefont
  {Bazavov}} \emph {et~al.} (\bibinfo {collaboration} {Fermilab Lattice,
  MILC}),\ }\bibfield  {title} {\bibinfo {title} {{Semileptonic form factors
  for $B \to D^\ast\ell\nu$ at nonzero recoil from 2 + 1-flavor lattice QCD}},\
  }\href@noop {} {\  (\bibinfo {year} {2021})},\ \Eprint
  {https://arxiv.org/abs/2105.14019} {arXiv:2105.14019 [hep-lat]} \BibitemShut
  {NoStop}%
\bibitem [{\citenamefont {Bordone}\ \emph {et~al.}(2021)\citenamefont
  {Bordone}, \citenamefont {Capdevila},\ and\ \citenamefont
  {Gambino}}]{Bordone:2021oof}%
  \BibitemOpen
  \bibfield  {author} {\bibinfo {author} {\bibfnamefont {M.}~\bibnamefont
  {Bordone}}, \bibinfo {author} {\bibfnamefont {B.}~\bibnamefont {Capdevila}},\
  and\ \bibinfo {author} {\bibfnamefont {P.}~\bibnamefont {Gambino}},\
  }\bibfield  {title} {\bibinfo {title} {{Three loop calculations and inclusive
  $V_{cb}$}},\ }\href {https://doi.org/10.1016/j.physletb.2021.136679}
  {\bibfield  {journal} {\bibinfo  {journal} {Phys. Lett. B}\ }\textbf
  {\bibinfo {volume} {822}},\ \bibinfo {pages} {136679} (\bibinfo {year}
  {2021})},\ \Eprint {https://arxiv.org/abs/2107.00604} {arXiv:2107.00604
  [hep-ph]} \BibitemShut {NoStop}%
\bibitem [{\citenamefont {Jay}\ \emph {et~al.}(2021)\citenamefont {Jay},
  \citenamefont {Lytle}, \citenamefont {DeTar}, \citenamefont {El-Khadra},
  \citenamefont {Gamiz}, \citenamefont {Gelzer}, \citenamefont {Gottlieb},
  \citenamefont {Kronfeld}, \citenamefont {Simone},\ and\ \citenamefont
  {Vaquero}}]{FermilabLattice:2021bxu}%
  \BibitemOpen
  \bibfield  {author} {\bibinfo {author} {\bibfnamefont {W.~I.}\ \bibnamefont
  {Jay}}, \bibinfo {author} {\bibfnamefont {A.}~\bibnamefont {Lytle}}, \bibinfo
  {author} {\bibfnamefont {C.}~\bibnamefont {DeTar}}, \bibinfo {author}
  {\bibfnamefont {A.}~\bibnamefont {El-Khadra}}, \bibinfo {author}
  {\bibfnamefont {E.}~\bibnamefont {Gamiz}}, \bibinfo {author} {\bibfnamefont
  {Z.}~\bibnamefont {Gelzer}}, \bibinfo {author} {\bibfnamefont
  {S.}~\bibnamefont {Gottlieb}}, \bibinfo {author} {\bibfnamefont
  {A.}~\bibnamefont {Kronfeld}}, \bibinfo {author} {\bibfnamefont
  {J.}~\bibnamefont {Simone}},\ and\ \bibinfo {author} {\bibfnamefont
  {A.}~\bibnamefont {Vaquero}} (\bibinfo {collaboration} {Fermilab Lattice,
  MILC}),\ }\bibfield  {title} {\bibinfo {title} {{B- and D-meson semileptonic
  decays with highly improved staggered quarks}}\ }(\bibinfo {year} {2021})\
  \Eprint {https://arxiv.org/abs/2111.05184} {arXiv:2111.05184 [hep-lat]}
  \BibitemShut {NoStop}%
\bibitem [{\citenamefont {Gonz\`alez-Sol\'\i{}s}\ \emph
  {et~al.}(2021)\citenamefont {Gonz\`alez-Sol\'\i{}s}, \citenamefont
  {Masjuan},\ and\ \citenamefont {Rojas}}]{Gonzalez-Solis:2021awb}%
  \BibitemOpen
  \bibfield  {author} {\bibinfo {author} {\bibfnamefont {S.}~\bibnamefont
  {Gonz\`alez-Sol\'\i{}s}}, \bibinfo {author} {\bibfnamefont {P.}~\bibnamefont
  {Masjuan}},\ and\ \bibinfo {author} {\bibfnamefont {C.}~\bibnamefont
  {Rojas}},\ }\bibfield  {title} {\bibinfo {title} {{Pad\'e approximants to
  $B\to\pi\ell\nu_{\ell}$ and $B_{s}\to K\ell\nu_{\ell}$ and determination of
  $|V_{ub}|$}},\ }\href@noop {} {\  (\bibinfo {year} {2021})},\ \Eprint
  {https://arxiv.org/abs/2110.06153} {arXiv:2110.06153 [hep-ph]} \BibitemShut
  {NoStop}%
\bibitem [{\citenamefont {Bansal}\ \emph {et~al.}(2021)\citenamefont {Bansal},
  \citenamefont {Mahajan},\ and\ \citenamefont {Mishra}}]{Bansal:2021oon}%
  \BibitemOpen
  \bibfield  {author} {\bibinfo {author} {\bibfnamefont {A.}~\bibnamefont
  {Bansal}}, \bibinfo {author} {\bibfnamefont {N.}~\bibnamefont {Mahajan}},\
  and\ \bibinfo {author} {\bibfnamefont {D.}~\bibnamefont {Mishra}},\
  }\bibfield  {title} {\bibinfo {title} {{$\frac{|V_{ub}|}{|V_{cb}|}$ and Quest
  for New Physics}},\ }\href@noop {} {\  (\bibinfo {year} {2021})},\ \Eprint
  {https://arxiv.org/abs/2112.00363} {arXiv:2112.00363 [hep-ph]} \BibitemShut
  {NoStop}%
\bibitem [{\citenamefont {Buras}(2003)}]{Buras:2003td}%
  \BibitemOpen
  \bibfield  {author} {\bibinfo {author} {\bibfnamefont {A.~J.}\ \bibnamefont
  {Buras}},\ }\bibfield  {title} {\bibinfo {title} {{Relations between $\Delta
  M_{s, d}$ and $B_{s,d} \to \mu \bar{\mu}$ in models with minimal flavor
  violation}},\ }\href {https://doi.org/10.1016/S0370-2693(03)00561-6}
  {\bibfield  {journal} {\bibinfo  {journal} {Phys. Lett. B}\ }\textbf
  {\bibinfo {volume} {566}},\ \bibinfo {pages} {115} (\bibinfo {year}
  {2003})},\ \Eprint {https://arxiv.org/abs/hep-ph/0303060}
  {arXiv:hep-ph/0303060} \BibitemShut {NoStop}%
\bibitem [{\citenamefont {Bobeth}\ and\ \citenamefont
  {Buras}(2021)}]{Bobeth:2021cxm}%
  \BibitemOpen
  \bibfield  {author} {\bibinfo {author} {\bibfnamefont {C.}~\bibnamefont
  {Bobeth}}\ and\ \bibinfo {author} {\bibfnamefont {A.~J.}\ \bibnamefont
  {Buras}},\ }\bibfield  {title} {\bibinfo {title} {{Searching for New Physics
  with $\overline{\mathcal{B}}(B_{s,d}\to\mu\bar\mu)/\Delta M_{s,d}$}}\ }\href
  {https://doi.org/10.5506/APhysPolB.52.1189} {10.5506/APhysPolB.52.1189}
  (\bibinfo {year} {2021}),\ \Eprint {https://arxiv.org/abs/2104.09521}
  {arXiv:2104.09521 [hep-ph]} \BibitemShut {NoStop}%
\bibitem [{\citenamefont {Buras}\ and\ \citenamefont
  {Venturini}(2021)}]{Buras:2021nns}%
  \BibitemOpen
  \bibfield  {author} {\bibinfo {author} {\bibfnamefont {A.~J.}\ \bibnamefont
  {Buras}}\ and\ \bibinfo {author} {\bibfnamefont {E.}~\bibnamefont
  {Venturini}},\ }\bibfield  {title} {\bibinfo {title} {{Searching for New
  Physics in Rare $K$ and $B$ Decays without $|V_{cb}|$ and $|V_{ub}|$
  Uncertainties}},\ }\href@noop {} {\  (\bibinfo {year} {2021})},\ \Eprint
  {https://arxiv.org/abs/2109.11032} {arXiv:2109.11032 [hep-ph]} \BibitemShut
  {NoStop}%
\bibitem [{\citenamefont {Brod}\ \emph {et~al.}(2020)\citenamefont {Brod},
  \citenamefont {Gorbahn},\ and\ \citenamefont {Stamou}}]{Brod:2019rzc}%
  \BibitemOpen
  \bibfield  {author} {\bibinfo {author} {\bibfnamefont {J.}~\bibnamefont
  {Brod}}, \bibinfo {author} {\bibfnamefont {M.}~\bibnamefont {Gorbahn}},\ and\
  \bibinfo {author} {\bibfnamefont {E.}~\bibnamefont {Stamou}},\ }\bibfield
  {title} {\bibinfo {title} {{Standard-Model Prediction of $\epsilon_K$ with
  Manifest Quark-Mixing Unitarity}},\ }\href
  {https://doi.org/10.1103/PhysRevLett.125.171803} {\bibfield  {journal}
  {\bibinfo  {journal} {Phys. Rev. Lett.}\ }\textbf {\bibinfo {volume} {125}},\
  \bibinfo {pages} {171803} (\bibinfo {year} {2020})},\ \Eprint
  {https://arxiv.org/abs/1911.06822} {arXiv:1911.06822 [hep-ph]} \BibitemShut
  {NoStop}%
\bibitem [{\citenamefont {Brod}\ \emph {et~al.}(2021)\citenamefont {Brod},
  \citenamefont {Kvedarait\.{e}},\ and\ \citenamefont
  {Polonsky}}]{Brod:2021qvc}%
  \BibitemOpen
  \bibfield  {author} {\bibinfo {author} {\bibfnamefont {J.}~\bibnamefont
  {Brod}}, \bibinfo {author} {\bibfnamefont {S.}~\bibnamefont
  {Kvedarait\.{e}}},\ and\ \bibinfo {author} {\bibfnamefont {Z.}~\bibnamefont
  {Polonsky}},\ }\bibfield  {title} {\bibinfo {title} {{Two-loop Electroweak
  Corrections to the Top-Quark Contribution to $\epsilon_K$}},\ }\href@noop {}
  {\  (\bibinfo {year} {2021})},\ \Eprint {https://arxiv.org/abs/2108.00017}
  {arXiv:2108.00017 [hep-ph]} \BibitemShut {NoStop}%
\bibitem [{\citenamefont {Aaij}\ \emph {et~al.}(2014)\citenamefont {Aaij} \emph
  {et~al.}}]{Aaij:2014pli}%
  \BibitemOpen
  \bibfield  {author} {\bibinfo {author} {\bibfnamefont {R.}~\bibnamefont
  {Aaij}} \emph {et~al.} (\bibinfo {collaboration} {LHCb}),\ }\bibfield
  {title} {\bibinfo {title} {{Differential branching fractions and isospin
  asymmetries of $B \to K^{(*)} \mu^+ \mu^-$ decays}},\ }\href
  {https://doi.org/10.1007/JHEP06(2014)133} {\bibfield  {journal} {\bibinfo
  {journal} {JHEP}\ }\textbf {\bibinfo {volume} {06}},\ \bibinfo {pages}
  {133}},\ \Eprint {https://arxiv.org/abs/1403.8044} {arXiv:1403.8044 [hep-ex]}
  \BibitemShut {NoStop}%
\bibitem [{\citenamefont {Aaij}\ \emph {et~al.}(2016)\citenamefont {Aaij} \emph
  {et~al.}}]{Aaij:2016flj}%
  \BibitemOpen
  \bibfield  {author} {\bibinfo {author} {\bibfnamefont {R.}~\bibnamefont
  {Aaij}} \emph {et~al.} (\bibinfo {collaboration} {LHCb}),\ }\bibfield
  {title} {\bibinfo {title} {{Measurements of the S-wave fraction in
  $B^{0}\rightarrow K^{+}\pi^{-}\mu^{+}\mu^{-}$ decays and the
  $B^{0}\rightarrow K^{\ast}(892)^{0}\mu^{+}\mu^{-}$ differential branching
  fraction}},\ }\href {https://doi.org/10.1007/JHEP11(2016)047} {\bibfield
  {journal} {\bibinfo  {journal} {JHEP}\ }\textbf {\bibinfo {volume} {11}},\
  \bibinfo {pages} {047}},\ \bibinfo {note} {[Erratum: JHEP 04, 142 (2017)]},\
  \Eprint {https://arxiv.org/abs/1606.04731} {arXiv:1606.04731 [hep-ex]}
  \BibitemShut {NoStop}%
\bibitem [{\citenamefont {Aaij}\ \emph
  {et~al.}(2021{\natexlab{a}})\citenamefont {Aaij} \emph
  {et~al.}}]{LHCb:2021zwz}%
  \BibitemOpen
  \bibfield  {author} {\bibinfo {author} {\bibfnamefont {R.}~\bibnamefont
  {Aaij}} \emph {et~al.} (\bibinfo {collaboration} {LHCb}),\ }\bibfield
  {title} {\bibinfo {title} {{Branching Fraction Measurements of the Rare
  $B^0_s\rightarrow\phi\mu^+\mu^-$ and $B^0_s\rightarrow
  f_2^\prime(1525)\mu^+\mu^-$ Decays}},\ }\href
  {https://doi.org/10.1103/PhysRevLett.127.151801} {\bibfield  {journal}
  {\bibinfo  {journal} {Phys. Rev. Lett.}\ }\textbf {\bibinfo {volume} {127}},\
  \bibinfo {pages} {151801} (\bibinfo {year} {2021}{\natexlab{a}})},\ \Eprint
  {https://arxiv.org/abs/2105.14007} {arXiv:2105.14007 [hep-ex]} \BibitemShut
  {NoStop}%
\bibitem [{\citenamefont {Altmannshofer}\ and\ \citenamefont
  {Straub}(2015)}]{Altmannshofer:2014rta}%
  \BibitemOpen
  \bibfield  {author} {\bibinfo {author} {\bibfnamefont {W.}~\bibnamefont
  {Altmannshofer}}\ and\ \bibinfo {author} {\bibfnamefont {D.~M.}\ \bibnamefont
  {Straub}},\ }\bibfield  {title} {\bibinfo {title} {{New physics in
  $b\rightarrow s$ transitions after LHC run 1}},\ }\href
  {https://doi.org/10.1140/epjc/s10052-015-3602-7} {\bibfield  {journal}
  {\bibinfo  {journal} {Eur. Phys. J. C}\ }\textbf {\bibinfo {volume} {75}},\
  \bibinfo {pages} {382} (\bibinfo {year} {2015})},\ \Eprint
  {https://arxiv.org/abs/1411.3161} {arXiv:1411.3161 [hep-ph]} \BibitemShut
  {NoStop}%
\bibitem [{\citenamefont {Cao}\ \emph {et~al.}(2021)\citenamefont {Cao} \emph
  {et~al.}}]{Belle:2021eni}%
  \BibitemOpen
  \bibfield  {author} {\bibinfo {author} {\bibfnamefont {L.}~\bibnamefont
  {Cao}} \emph {et~al.} (\bibinfo {collaboration} {Belle}),\ }\bibfield
  {title} {\bibinfo {title} {{Measurements of Partial Branching Fractions of
  Inclusive $B \to X_u \, \ell^+\, \nu_{\ell}$ Decays with Hadronic Tagging}},\
  }\href {https://doi.org/10.1103/PhysRevD.104.012008} {\bibfield  {journal}
  {\bibinfo  {journal} {Phys. Rev. D}\ }\textbf {\bibinfo {volume} {104}},\
  \bibinfo {pages} {012008} (\bibinfo {year} {2021})},\ \Eprint
  {https://arxiv.org/abs/2102.00020} {arXiv:2102.00020 [hep-ex]} \BibitemShut
  {NoStop}%
\bibitem [{\citenamefont {Waheed}\ \emph {et~al.}(2019)\citenamefont {Waheed}
  \emph {et~al.}}]{Belle:2018ezy}%
  \BibitemOpen
  \bibfield  {author} {\bibinfo {author} {\bibfnamefont {E.}~\bibnamefont
  {Waheed}} \emph {et~al.} (\bibinfo {collaboration} {Belle}),\ }\bibfield
  {title} {\bibinfo {title} {{Measurement of the CKM matrix element $|V_{cb}|$
  from $B^0\to D^{*-}\ell^ {+} \nu_\ell$ at Belle}},\ }\href
  {https://doi.org/10.1103/PhysRevD.100.052007} {\bibfield  {journal} {\bibinfo
   {journal} {Phys. Rev. D}\ }\textbf {\bibinfo {volume} {100}},\ \bibinfo
  {pages} {052007} (\bibinfo {year} {2019})},\ \bibinfo {note} {[Erratum:
  Phys.Rev.D 103, 079901 (2021)]},\ \Eprint {https://arxiv.org/abs/1809.03290}
  {arXiv:1809.03290 [hep-ex]} \BibitemShut {NoStop}%
\bibitem [{\citenamefont {Lees}\ \emph {et~al.}(2019)\citenamefont {Lees} \emph
  {et~al.}}]{BaBar:2019vpl}%
  \BibitemOpen
  \bibfield  {author} {\bibinfo {author} {\bibfnamefont {J.~P.}\ \bibnamefont
  {Lees}} \emph {et~al.} (\bibinfo {collaboration} {BaBar}),\ }\bibfield
  {title} {\bibinfo {title} {{Extraction of form Factors from a
  Four-Dimensional Angular Analysis of $\overline{B} \rightarrow D^\ast \ell^-
  \overline{\nu}_\ell$}},\ }\href
  {https://doi.org/10.1103/PhysRevLett.123.091801} {\bibfield  {journal}
  {\bibinfo  {journal} {Phys. Rev. Lett.}\ }\textbf {\bibinfo {volume} {123}},\
  \bibinfo {pages} {091801} (\bibinfo {year} {2019})},\ \Eprint
  {https://arxiv.org/abs/1903.10002} {arXiv:1903.10002 [hep-ex]} \BibitemShut
  {NoStop}%
\bibitem [{\citenamefont {Aaij}\ \emph
  {et~al.}(2015{\natexlab{a}})\citenamefont {Aaij} \emph
  {et~al.}}]{Aaij:2015bfa}%
  \BibitemOpen
  \bibfield  {author} {\bibinfo {author} {\bibfnamefont {R.}~\bibnamefont
  {Aaij}} \emph {et~al.} (\bibinfo {collaboration} {LHCb}),\ }\bibfield
  {title} {\bibinfo {title} {{Determination of the quark coupling strength
  $|V_{ub}|$ using baryonic decays}},\ }\href
  {https://doi.org/10.1038/nphys3415} {\bibfield  {journal} {\bibinfo
  {journal} {Nature Phys.}\ }\textbf {\bibinfo {volume} {11}},\ \bibinfo
  {pages} {743} (\bibinfo {year} {2015}{\natexlab{a}})},\ \Eprint
  {https://arxiv.org/abs/1504.01568} {arXiv:1504.01568 [hep-ex]} \BibitemShut
  {NoStop}%
\bibitem [{\citenamefont {Detmold}\ \emph {et~al.}(2015)\citenamefont
  {Detmold}, \citenamefont {Lehner},\ and\ \citenamefont
  {Meinel}}]{Detmold:2015aaa}%
  \BibitemOpen
  \bibfield  {author} {\bibinfo {author} {\bibfnamefont {W.}~\bibnamefont
  {Detmold}}, \bibinfo {author} {\bibfnamefont {C.}~\bibnamefont {Lehner}},\
  and\ \bibinfo {author} {\bibfnamefont {S.}~\bibnamefont {Meinel}},\
  }\bibfield  {title} {\bibinfo {title} {{$\Lambda_b \to p \ell^-
  \bar{\nu}_\ell$ and $\Lambda_b \to \Lambda_c \ell^- \bar{\nu}_\ell$ form
  factors from lattice QCD with relativistic heavy quarks}},\ }\href
  {https://doi.org/10.1103/PhysRevD.92.034503} {\bibfield  {journal} {\bibinfo
  {journal} {Phys. Rev. D}\ }\textbf {\bibinfo {volume} {92}},\ \bibinfo
  {pages} {034503} (\bibinfo {year} {2015})},\ \Eprint
  {https://arxiv.org/abs/1503.01421} {arXiv:1503.01421 [hep-lat]} \BibitemShut
  {NoStop}%
\bibitem [{\citenamefont {Amhis}\ \emph {et~al.}(2019)\citenamefont {Amhis}
  \emph {et~al.}}]{Amhis:2019ckw}%
  \BibitemOpen
  \bibfield  {author} {\bibinfo {author} {\bibfnamefont {Y.~S.}\ \bibnamefont
  {Amhis}} \emph {et~al.} (\bibinfo {collaboration} {HFLAV}),\ }\bibfield
  {title} {\bibinfo {title} {{Averages of $b$-hadron, $c$-hadron, and
  $\tau$-lepton properties as of 2018}},\ }\href@noop {} {\  (\bibinfo {year}
  {2019})},\ \Eprint {https://arxiv.org/abs/1909.12524} {arXiv:1909.12524
  [hep-ex]} \BibitemShut {NoStop}%
\bibitem [{\citenamefont {Aaij}\ \emph
  {et~al.}(2021{\natexlab{b}})\citenamefont {Aaij} \emph
  {et~al.}}]{LHCb:2020ist}%
  \BibitemOpen
  \bibfield  {author} {\bibinfo {author} {\bibfnamefont {R.}~\bibnamefont
  {Aaij}} \emph {et~al.} (\bibinfo {collaboration} {LHCb}),\ }\bibfield
  {title} {\bibinfo {title} {{First observation of the decay $B_s^0 \to
  K^-\mu^+\nu_\mu$ and Measurement of $|V_{ub}|/|V_{cb}|$}},\ }\href
  {https://doi.org/10.1103/PhysRevLett.126.081804} {\bibfield  {journal}
  {\bibinfo  {journal} {Phys. Rev. Lett.}\ }\textbf {\bibinfo {volume} {126}},\
  \bibinfo {pages} {081804} (\bibinfo {year} {2021}{\natexlab{b}})},\ \Eprint
  {https://arxiv.org/abs/2012.05143} {arXiv:2012.05143 [hep-ex]} \BibitemShut
  {NoStop}%
\bibitem [{\citenamefont {Bouchard}\ \emph {et~al.}(2014)\citenamefont
  {Bouchard}, \citenamefont {Lepage}, \citenamefont {Monahan}, \citenamefont
  {Na},\ and\ \citenamefont {Shigemitsu}}]{Bouchard:2014ypa}%
  \BibitemOpen
  \bibfield  {author} {\bibinfo {author} {\bibfnamefont {C.~M.}\ \bibnamefont
  {Bouchard}}, \bibinfo {author} {\bibfnamefont {G.~P.}\ \bibnamefont
  {Lepage}}, \bibinfo {author} {\bibfnamefont {C.}~\bibnamefont {Monahan}},
  \bibinfo {author} {\bibfnamefont {H.}~\bibnamefont {Na}},\ and\ \bibinfo
  {author} {\bibfnamefont {J.}~\bibnamefont {Shigemitsu}},\ }\bibfield  {title}
  {\bibinfo {title} {{$B_s \to K \ell \nu$ form factors from lattice QCD}},\
  }\href {https://doi.org/10.1103/PhysRevD.90.054506} {\bibfield  {journal}
  {\bibinfo  {journal} {Phys. Rev. D}\ }\textbf {\bibinfo {volume} {90}},\
  \bibinfo {pages} {054506} (\bibinfo {year} {2014})},\ \Eprint
  {https://arxiv.org/abs/1406.2279} {arXiv:1406.2279 [hep-lat]} \BibitemShut
  {NoStop}%
\bibitem [{\citenamefont {Flynn}\ \emph {et~al.}(2015)\citenamefont {Flynn},
  \citenamefont {Izubuchi}, \citenamefont {Kawanai}, \citenamefont {Lehner},
  \citenamefont {Soni}, \citenamefont {Van~de Water},\ and\ \citenamefont
  {Witzel}}]{Flynn:2015mha}%
  \BibitemOpen
  \bibfield  {author} {\bibinfo {author} {\bibfnamefont {J.~M.}\ \bibnamefont
  {Flynn}}, \bibinfo {author} {\bibfnamefont {T.}~\bibnamefont {Izubuchi}},
  \bibinfo {author} {\bibfnamefont {T.}~\bibnamefont {Kawanai}}, \bibinfo
  {author} {\bibfnamefont {C.}~\bibnamefont {Lehner}}, \bibinfo {author}
  {\bibfnamefont {A.}~\bibnamefont {Soni}}, \bibinfo {author} {\bibfnamefont
  {R.~S.}\ \bibnamefont {Van~de Water}},\ and\ \bibinfo {author} {\bibfnamefont
  {O.}~\bibnamefont {Witzel}},\ }\bibfield  {title} {\bibinfo {title} {{$B \to
  \pi \ell \nu$ and $B_s \to K \ell \nu$ form factors and $|V_{ub}|$ from
  2+1-flavor lattice QCD with domain-wall light quarks and relativistic heavy
  quarks}},\ }\href {https://doi.org/10.1103/PhysRevD.91.074510} {\bibfield
  {journal} {\bibinfo  {journal} {Phys. Rev. D}\ }\textbf {\bibinfo {volume}
  {91}},\ \bibinfo {pages} {074510} (\bibinfo {year} {2015})},\ \Eprint
  {https://arxiv.org/abs/1501.05373} {arXiv:1501.05373 [hep-lat]} \BibitemShut
  {NoStop}%
\bibitem [{\citenamefont {Bazavov}\ \emph {et~al.}(2019)\citenamefont {Bazavov}
  \emph {et~al.}}]{FermilabLattice:2019ikx}%
  \BibitemOpen
  \bibfield  {author} {\bibinfo {author} {\bibfnamefont {A.}~\bibnamefont
  {Bazavov}} \emph {et~al.} (\bibinfo {collaboration} {Fermilab Lattice,
  MILC}),\ }\bibfield  {title} {\bibinfo {title} {{$B_s\to K\ell\nu$ decay from
  lattice QCD}},\ }\href {https://doi.org/10.1103/PhysRevD.100.034501}
  {\bibfield  {journal} {\bibinfo  {journal} {Phys. Rev. D}\ }\textbf {\bibinfo
  {volume} {100}},\ \bibinfo {pages} {034501} (\bibinfo {year} {2019})},\
  \Eprint {https://arxiv.org/abs/1901.02561} {arXiv:1901.02561 [hep-lat]}
  \BibitemShut {NoStop}%
\bibitem [{\citenamefont {McLean}\ \emph {et~al.}(2020)\citenamefont {McLean},
  \citenamefont {Davies}, \citenamefont {Koponen},\ and\ \citenamefont
  {Lytle}}]{McLean:2019qcx}%
  \BibitemOpen
  \bibfield  {author} {\bibinfo {author} {\bibfnamefont {E.}~\bibnamefont
  {McLean}}, \bibinfo {author} {\bibfnamefont {C.~T.~H.}\ \bibnamefont
  {Davies}}, \bibinfo {author} {\bibfnamefont {J.}~\bibnamefont {Koponen}},\
  and\ \bibinfo {author} {\bibfnamefont {A.~T.}\ \bibnamefont {Lytle}},\
  }\bibfield  {title} {\bibinfo {title} {{$B_s\to D_s \ell\nu$ Form Factors for
  the full $q^2$ range from Lattice QCD with non-perturbatively normalized
  currents}},\ }\href {https://doi.org/10.1103/PhysRevD.101.074513} {\bibfield
  {journal} {\bibinfo  {journal} {Phys. Rev. D}\ }\textbf {\bibinfo {volume}
  {101}},\ \bibinfo {pages} {074513} (\bibinfo {year} {2020})},\ \Eprint
  {https://arxiv.org/abs/1906.00701} {arXiv:1906.00701 [hep-lat]} \BibitemShut
  {NoStop}%
\bibitem [{\citenamefont {Aoki}\ \emph {et~al.}(2021)\citenamefont {Aoki} \emph
  {et~al.}}]{Aoki:2021kgd}%
  \BibitemOpen
  \bibfield  {author} {\bibinfo {author} {\bibfnamefont {Y.}~\bibnamefont
  {Aoki}} \emph {et~al.},\ }\bibfield  {title} {\bibinfo {title} {{FLAG Review
  2021}},\ }\href@noop {} {\  (\bibinfo {year} {2021})},\ \Eprint
  {https://arxiv.org/abs/2111.09849} {arXiv:2111.09849 [hep-lat]} \BibitemShut
  {NoStop}%
\bibitem [{\citenamefont {Khodjamirian}\ and\ \citenamefont
  {Rusov}(2017)}]{Khodjamirian:2017fxg}%
  \BibitemOpen
  \bibfield  {author} {\bibinfo {author} {\bibfnamefont {A.}~\bibnamefont
  {Khodjamirian}}\ and\ \bibinfo {author} {\bibfnamefont {A.~V.}\ \bibnamefont
  {Rusov}},\ }\bibfield  {title} {\bibinfo {title} {{$B_{s}\to K \ell \nu_\ell$
  and $B_{(s)} \to \pi (K) \ell^+\ell^-$ decays at large recoil and CKM matrix
  elements}},\ }\href {https://doi.org/10.1007/JHEP08(2017)112} {\bibfield
  {journal} {\bibinfo  {journal} {JHEP}\ }\textbf {\bibinfo {volume} {08}},\
  \bibinfo {pages} {112}},\ \Eprint {https://arxiv.org/abs/1703.04765}
  {arXiv:1703.04765 [hep-ph]} \BibitemShut {NoStop}%
\bibitem [{\citenamefont {Lees}\ \emph {et~al.}(2013)\citenamefont {Lees} \emph
  {et~al.}}]{BaBar:2012nus}%
  \BibitemOpen
  \bibfield  {author} {\bibinfo {author} {\bibfnamefont {J.~P.}\ \bibnamefont
  {Lees}} \emph {et~al.} (\bibinfo {collaboration} {BaBar}),\ }\bibfield
  {title} {\bibinfo {title} {{Evidence of $B^+ \to \tau^+\nu$ decays with
  hadronic B tags}},\ }\href {https://doi.org/10.1103/PhysRevD.88.031102}
  {\bibfield  {journal} {\bibinfo  {journal} {Phys. Rev. D}\ }\textbf {\bibinfo
  {volume} {88}},\ \bibinfo {pages} {031102} (\bibinfo {year} {2013})},\
  \Eprint {https://arxiv.org/abs/1207.0698} {arXiv:1207.0698 [hep-ex]}
  \BibitemShut {NoStop}%
\bibitem [{\citenamefont {Kronenbitter}\ \emph {et~al.}(2015)\citenamefont
  {Kronenbitter} \emph {et~al.}}]{Belle:2015odw}%
  \BibitemOpen
  \bibfield  {author} {\bibinfo {author} {\bibfnamefont {B.}~\bibnamefont
  {Kronenbitter}} \emph {et~al.} (\bibinfo {collaboration} {Belle}),\
  }\bibfield  {title} {\bibinfo {title} {{Measurement of the branching fraction
  of $B^+ \to \tau^+ \nu_\tau$ decays with the semileptonic tagging method}},\
  }\href {https://doi.org/10.1103/PhysRevD.92.051102} {\bibfield  {journal}
  {\bibinfo  {journal} {Phys. Rev. D}\ }\textbf {\bibinfo {volume} {92}},\
  \bibinfo {pages} {051102} (\bibinfo {year} {2015})},\ \Eprint
  {https://arxiv.org/abs/1503.05613} {arXiv:1503.05613 [hep-ex]} \BibitemShut
  {NoStop}%
\bibitem [{\citenamefont {Wolfenstein}(1983)}]{Wolfenstein:1983yz}%
  \BibitemOpen
  \bibfield  {author} {\bibinfo {author} {\bibfnamefont {L.}~\bibnamefont
  {Wolfenstein}},\ }\bibfield  {title} {\bibinfo {title} {{Parametrization of
  the Kobayashi-Maskawa Matrix}},\ }\href
  {https://doi.org/10.1103/PhysRevLett.51.1945} {\bibfield  {journal} {\bibinfo
   {journal} {Phys. Rev. Lett.}\ }\textbf {\bibinfo {volume} {51}},\ \bibinfo
  {pages} {1945} (\bibinfo {year} {1983})}\BibitemShut {NoStop}%
\bibitem [{\citenamefont {Buras}\ and\ \citenamefont
  {Guadagnoli}(2008)}]{Buras:2008nn}%
  \BibitemOpen
  \bibfield  {author} {\bibinfo {author} {\bibfnamefont {A.~J.}\ \bibnamefont
  {Buras}}\ and\ \bibinfo {author} {\bibfnamefont {D.}~\bibnamefont
  {Guadagnoli}},\ }\bibfield  {title} {\bibinfo {title} {{Correlations among
  new CP violating effects in $\Delta F = 2$ observables}},\ }\href
  {https://doi.org/10.1103/PhysRevD.78.033005} {\bibfield  {journal} {\bibinfo
  {journal} {Phys. Rev. D}\ }\textbf {\bibinfo {volume} {78}},\ \bibinfo
  {pages} {033005} (\bibinfo {year} {2008})},\ \Eprint
  {https://arxiv.org/abs/0805.3887} {arXiv:0805.3887 [hep-ph]} \BibitemShut
  {NoStop}%
\bibitem [{\citenamefont {Aaij}\ \emph
  {et~al.}(2021{\natexlab{c}})\citenamefont {Aaij} \emph
  {et~al.}}]{LHCb:2021moh}%
  \BibitemOpen
  \bibfield  {author} {\bibinfo {author} {\bibfnamefont {R.}~\bibnamefont
  {Aaij}} \emph {et~al.} (\bibinfo {collaboration} {LHCb}),\ }\bibfield
  {title} {\bibinfo {title} {{Precise determination of the
  $B^0_s$-$\overline{B}^0_s$ oscillation frequency}},\ }\href@noop {} {\
  (\bibinfo {year} {2021}{\natexlab{c}})},\ \Eprint
  {https://arxiv.org/abs/2104.04421} {arXiv:2104.04421 [hep-ex]} \BibitemShut
  {NoStop}%
\bibitem [{\citenamefont {Dowdall}\ \emph {et~al.}(2019)\citenamefont
  {Dowdall}, \citenamefont {Davies}, \citenamefont {Horgan}, \citenamefont
  {Lepage}, \citenamefont {Monahan}, \citenamefont {Shigemitsu},\ and\
  \citenamefont {Wingate}}]{Dowdall:2019bea}%
  \BibitemOpen
  \bibfield  {author} {\bibinfo {author} {\bibfnamefont {R.~J.}\ \bibnamefont
  {Dowdall}}, \bibinfo {author} {\bibfnamefont {C.~T.~H.}\ \bibnamefont
  {Davies}}, \bibinfo {author} {\bibfnamefont {R.~R.}\ \bibnamefont {Horgan}},
  \bibinfo {author} {\bibfnamefont {G.~P.}\ \bibnamefont {Lepage}}, \bibinfo
  {author} {\bibfnamefont {C.~J.}\ \bibnamefont {Monahan}}, \bibinfo {author}
  {\bibfnamefont {J.}~\bibnamefont {Shigemitsu}},\ and\ \bibinfo {author}
  {\bibfnamefont {M.}~\bibnamefont {Wingate}},\ }\bibfield  {title} {\bibinfo
  {title} {{Neutral B-meson mixing from full lattice QCD at the physical
  point}},\ }\href {https://doi.org/10.1103/PhysRevD.100.094508} {\bibfield
  {journal} {\bibinfo  {journal} {Phys. Rev. D}\ }\textbf {\bibinfo {volume}
  {100}},\ \bibinfo {pages} {094508} (\bibinfo {year} {2019})},\ \Eprint
  {https://arxiv.org/abs/1907.01025} {arXiv:1907.01025 [hep-lat]} \BibitemShut
  {NoStop}%
\bibitem [{\citenamefont {King}\ \emph
  {et~al.}(2019{\natexlab{a}})\citenamefont {King}, \citenamefont {Lenz},\ and\
  \citenamefont {Rauh}}]{King:2019lal}%
  \BibitemOpen
  \bibfield  {author} {\bibinfo {author} {\bibfnamefont {D.}~\bibnamefont
  {King}}, \bibinfo {author} {\bibfnamefont {A.}~\bibnamefont {Lenz}},\ and\
  \bibinfo {author} {\bibfnamefont {T.}~\bibnamefont {Rauh}},\ }\bibfield
  {title} {\bibinfo {title} {{$B_{s}$ mixing observables and $|V_{td}/V_{ts}|$
  from sum rules}},\ }\href {https://doi.org/10.1007/JHEP05(2019)034}
  {\bibfield  {journal} {\bibinfo  {journal} {JHEP}\ }\textbf {\bibinfo
  {volume} {05}},\ \bibinfo {pages} {034}},\ \Eprint
  {https://arxiv.org/abs/1904.00940} {arXiv:1904.00940 [hep-ph]} \BibitemShut
  {NoStop}%
\bibitem [{\citenamefont {Di~Luzio}\ \emph {et~al.}(2018)\citenamefont
  {Di~Luzio}, \citenamefont {Kirk},\ and\ \citenamefont
  {Lenz}}]{DiLuzio:2017fdq}%
  \BibitemOpen
  \bibfield  {author} {\bibinfo {author} {\bibfnamefont {L.}~\bibnamefont
  {Di~Luzio}}, \bibinfo {author} {\bibfnamefont {M.}~\bibnamefont {Kirk}},\
  and\ \bibinfo {author} {\bibfnamefont {A.}~\bibnamefont {Lenz}},\ }\bibfield
  {title} {\bibinfo {title} {{Updated $B_s$-mixing constraints on new physics
  models for $b\to s\ell^+\ell^-$ anomalies}},\ }\href
  {https://doi.org/10.1103/PhysRevD.97.095035} {\bibfield  {journal} {\bibinfo
  {journal} {Phys. Rev. D}\ }\textbf {\bibinfo {volume} {97}},\ \bibinfo
  {pages} {095035} (\bibinfo {year} {2018})},\ \Eprint
  {https://arxiv.org/abs/1712.06572} {arXiv:1712.06572 [hep-ph]} \BibitemShut
  {NoStop}%
\bibitem [{\citenamefont {Bernlochner}\ \emph {et~al.}(2021)\citenamefont
  {Bernlochner}, \citenamefont {Lacker}, \citenamefont {Ligeti}, \citenamefont
  {Stewart}, \citenamefont {Tackmann},\ and\ \citenamefont
  {Tackmann}}]{Bernlochner:2020jlt}%
  \BibitemOpen
  \bibfield  {author} {\bibinfo {author} {\bibfnamefont {F.~U.}\ \bibnamefont
  {Bernlochner}}, \bibinfo {author} {\bibfnamefont {H.}~\bibnamefont {Lacker}},
  \bibinfo {author} {\bibfnamefont {Z.}~\bibnamefont {Ligeti}}, \bibinfo
  {author} {\bibfnamefont {I.~W.}\ \bibnamefont {Stewart}}, \bibinfo {author}
  {\bibfnamefont {F.~J.}\ \bibnamefont {Tackmann}},\ and\ \bibinfo {author}
  {\bibfnamefont {K.}~\bibnamefont {Tackmann}} (\bibinfo {collaboration}
  {SIMBA}),\ }\bibfield  {title} {\bibinfo {title} {{Precision Global
  Determination of the $B\to X_s \gamma$ Decay Rate}},\ }\href
  {https://doi.org/10.1103/PhysRevLett.127.102001} {\bibfield  {journal}
  {\bibinfo  {journal} {Phys. Rev. Lett.}\ }\textbf {\bibinfo {volume} {127}},\
  \bibinfo {pages} {102001} (\bibinfo {year} {2021})},\ \Eprint
  {https://arxiv.org/abs/2007.04320} {arXiv:2007.04320 [hep-ph]} \BibitemShut
  {NoStop}%
\bibitem [{\citenamefont {Bazavov}\ \emph {et~al.}(2018)\citenamefont {Bazavov}
  \emph {et~al.}}]{Bazavov:2017lyh}%
  \BibitemOpen
  \bibfield  {author} {\bibinfo {author} {\bibfnamefont {A.}~\bibnamefont
  {Bazavov}} \emph {et~al.},\ }\bibfield  {title} {\bibinfo {title} {{$B$- and
  $D$-meson leptonic decay constants from four-flavor lattice QCD}},\ }\href
  {https://doi.org/10.1103/PhysRevD.98.074512} {\bibfield  {journal} {\bibinfo
  {journal} {Phys. Rev. D}\ }\textbf {\bibinfo {volume} {98}},\ \bibinfo
  {pages} {074512} (\bibinfo {year} {2018})},\ \Eprint
  {https://arxiv.org/abs/1712.09262} {arXiv:1712.09262 [hep-lat]} \BibitemShut
  {NoStop}%
\bibitem [{\citenamefont {Altmannshofer}\ and\ \citenamefont
  {Stangl}(2021)}]{Altmannshofer:2021qrr}%
  \BibitemOpen
  \bibfield  {author} {\bibinfo {author} {\bibfnamefont {W.}~\bibnamefont
  {Altmannshofer}}\ and\ \bibinfo {author} {\bibfnamefont {P.}~\bibnamefont
  {Stangl}},\ }\bibfield  {title} {\bibinfo {title} {{New physics in rare B
  decays after Moriond 2021}},\ }\href
  {https://doi.org/10.1140/epjc/s10052-021-09725-1} {\bibfield  {journal}
  {\bibinfo  {journal} {Eur. Phys. J. C}\ }\textbf {\bibinfo {volume} {81}},\
  \bibinfo {pages} {952} (\bibinfo {year} {2021})},\ \Eprint
  {https://arxiv.org/abs/2103.13370} {arXiv:2103.13370 [hep-ph]} \BibitemShut
  {NoStop}%
\bibitem [{\citenamefont {Aaij}\ \emph
  {et~al.}(2021{\natexlab{d}})\citenamefont {Aaij} \emph
  {et~al.}}]{LHCb:2021awg}%
  \BibitemOpen
  \bibfield  {author} {\bibinfo {author} {\bibfnamefont {R.}~\bibnamefont
  {Aaij}} \emph {et~al.} (\bibinfo {collaboration} {LHCb}),\ }\bibfield
  {title} {\bibinfo {title} {{Measurement of the $B^0_s\to\mu^+\mu^-$ decay
  properties and search for the $B^0\to\mu^+\mu^-$ and
  $B^0_s\to\mu^+\mu^-\gamma$ decays}},\ }\href@noop {} {\  (\bibinfo {year}
  {2021}{\natexlab{d}})},\ \Eprint {https://arxiv.org/abs/2108.09283}
  {arXiv:2108.09283 [hep-ex]} \BibitemShut {NoStop}%
\bibitem [{\citenamefont {Aaij}\ \emph
  {et~al.}(2021{\natexlab{e}})\citenamefont {Aaij} \emph
  {et~al.}}]{LHCb:2021vsc}%
  \BibitemOpen
  \bibfield  {author} {\bibinfo {author} {\bibfnamefont {R.}~\bibnamefont
  {Aaij}} \emph {et~al.} (\bibinfo {collaboration} {LHCb}),\ }\bibfield
  {title} {\bibinfo {title} {{Analysis of neutral $B$-meson decays into two
  muons}},\ }\href@noop {} {\  (\bibinfo {year} {2021}{\natexlab{e}})},\
  \Eprint {https://arxiv.org/abs/2108.09284} {arXiv:2108.09284 [hep-ex]}
  \BibitemShut {NoStop}%
\bibitem [{\citenamefont {Aaboud}\ \emph {et~al.}(2019)\citenamefont {Aaboud}
  \emph {et~al.}}]{Aaboud:2018mst}%
  \BibitemOpen
  \bibfield  {author} {\bibinfo {author} {\bibfnamefont {M.}~\bibnamefont
  {Aaboud}} \emph {et~al.} (\bibinfo {collaboration} {ATLAS}),\ }\bibfield
  {title} {\bibinfo {title} {{Study of the rare decays of $B^0_s$ and $B^0$
  mesons into muon pairs using data collected during 2015 and 2016 with the
  ATLAS detector}},\ }\href {https://doi.org/10.1007/JHEP04(2019)098}
  {\bibfield  {journal} {\bibinfo  {journal} {JHEP}\ }\textbf {\bibinfo
  {volume} {04}},\ \bibinfo {pages} {098}},\ \Eprint
  {https://arxiv.org/abs/1812.03017} {arXiv:1812.03017 [hep-ex]} \BibitemShut
  {NoStop}%
\bibitem [{\citenamefont {Sirunyan}\ \emph {et~al.}(2020)\citenamefont
  {Sirunyan} \emph {et~al.}}]{Sirunyan:2019xdu}%
  \BibitemOpen
  \bibfield  {author} {\bibinfo {author} {\bibfnamefont {A.~M.}\ \bibnamefont
  {Sirunyan}} \emph {et~al.} (\bibinfo {collaboration} {CMS}),\ }\bibfield
  {title} {\bibinfo {title} {{Measurement of properties of
  B$^0_\mathrm{s}\to\mu^+\mu^-$ decays and search for B$^0\to\mu^+\mu^-$ with
  the CMS experiment}},\ }\href {https://doi.org/10.1007/JHEP04(2020)188}
  {\bibfield  {journal} {\bibinfo  {journal} {JHEP}\ }\textbf {\bibinfo
  {volume} {04}},\ \bibinfo {pages} {188}},\ \Eprint
  {https://arxiv.org/abs/1910.12127} {arXiv:1910.12127 [hep-ex]} \BibitemShut
  {NoStop}%
\bibitem [{\citenamefont {Beneke}\ \emph {et~al.}(2019)\citenamefont {Beneke},
  \citenamefont {Bobeth},\ and\ \citenamefont {Szafron}}]{Beneke:2019slt}%
  \BibitemOpen
  \bibfield  {author} {\bibinfo {author} {\bibfnamefont {M.}~\bibnamefont
  {Beneke}}, \bibinfo {author} {\bibfnamefont {C.}~\bibnamefont {Bobeth}},\
  and\ \bibinfo {author} {\bibfnamefont {R.}~\bibnamefont {Szafron}},\
  }\bibfield  {title} {\bibinfo {title} {{Power-enhanced leading-logarithmic
  QED corrections to $B_q \to \mu^+\mu^-$}},\ }\href
  {https://doi.org/10.1007/JHEP10(2019)232} {\bibfield  {journal} {\bibinfo
  {journal} {JHEP}\ }\textbf {\bibinfo {volume} {10}},\ \bibinfo {pages}
  {232}},\ \Eprint {https://arxiv.org/abs/1908.07011} {arXiv:1908.07011
  [hep-ph]} \BibitemShut {NoStop}%
\bibitem [{\citenamefont {Straub}(2018)}]{Straub:2018kue}%
  \BibitemOpen
  \bibfield  {author} {\bibinfo {author} {\bibfnamefont {D.~M.}\ \bibnamefont
  {Straub}},\ }\bibfield  {title} {\bibinfo {title} {{flavio: a Python package
  for flavour and precision phenomenology in the Standard Model and beyond}},\
  }\href@noop {} {\  (\bibinfo {year} {2018})},\ \Eprint
  {https://arxiv.org/abs/1810.08132} {arXiv:1810.08132 [hep-ph]} \BibitemShut
  {NoStop}%
\bibitem [{\citenamefont {Bharucha}\ \emph {et~al.}(2016)\citenamefont
  {Bharucha}, \citenamefont {Straub},\ and\ \citenamefont
  {Zwicky}}]{Bharucha:2015bzk}%
  \BibitemOpen
  \bibfield  {author} {\bibinfo {author} {\bibfnamefont {A.}~\bibnamefont
  {Bharucha}}, \bibinfo {author} {\bibfnamefont {D.~M.}\ \bibnamefont
  {Straub}},\ and\ \bibinfo {author} {\bibfnamefont {R.}~\bibnamefont
  {Zwicky}},\ }\bibfield  {title} {\bibinfo {title} {{$B\to V\ell^+\ell^-$ in
  the Standard Model from light-cone sum rules}},\ }\href
  {https://doi.org/10.1007/JHEP08(2016)098} {\bibfield  {journal} {\bibinfo
  {journal} {JHEP}\ }\textbf {\bibinfo {volume} {08}},\ \bibinfo {pages}
  {098}},\ \Eprint {https://arxiv.org/abs/1503.05534} {arXiv:1503.05534
  [hep-ph]} \BibitemShut {NoStop}%
\bibitem [{\citenamefont {Gubernari}\ \emph {et~al.}(2019)\citenamefont
  {Gubernari}, \citenamefont {Kokulu},\ and\ \citenamefont {van
  Dyk}}]{Gubernari:2018wyi}%
  \BibitemOpen
  \bibfield  {author} {\bibinfo {author} {\bibfnamefont {N.}~\bibnamefont
  {Gubernari}}, \bibinfo {author} {\bibfnamefont {A.}~\bibnamefont {Kokulu}},\
  and\ \bibinfo {author} {\bibfnamefont {D.}~\bibnamefont {van Dyk}},\
  }\bibfield  {title} {\bibinfo {title} {{$B\to P$ and $B\to V$ Form Factors
  from $B$-Meson Light-Cone Sum Rules beyond Leading Twist}},\ }\href
  {https://doi.org/10.1007/JHEP01(2019)150} {\bibfield  {journal} {\bibinfo
  {journal} {JHEP}\ }\textbf {\bibinfo {volume} {01}},\ \bibinfo {pages}
  {150}},\ \Eprint {https://arxiv.org/abs/1811.00983} {arXiv:1811.00983
  [hep-ph]} \BibitemShut {NoStop}%
\bibitem [{\citenamefont {Aaij}\ \emph
  {et~al.}(2015{\natexlab{b}})\citenamefont {Aaij} \emph
  {et~al.}}]{LHCb:2015tgy}%
  \BibitemOpen
  \bibfield  {author} {\bibinfo {author} {\bibfnamefont {R.}~\bibnamefont
  {Aaij}} \emph {et~al.} (\bibinfo {collaboration} {LHCb}),\ }\bibfield
  {title} {\bibinfo {title} {{Differential branching fraction and angular
  analysis of $\Lambda^{0}_{b} \rightarrow \Lambda \mu^+\mu^-$ decays}},\
  }\href {https://doi.org/10.1007/JHEP06(2015)115} {\bibfield  {journal}
  {\bibinfo  {journal} {JHEP}\ }\textbf {\bibinfo {volume} {06}},\ \bibinfo
  {pages} {115}},\ \bibinfo {note} {[Erratum: JHEP 09, 145 (2018)]},\ \Eprint
  {https://arxiv.org/abs/1503.07138} {arXiv:1503.07138 [hep-ex]} \BibitemShut
  {NoStop}%
\bibitem [{\citenamefont {Charles}\ \emph {et~al.}(2005)\citenamefont
  {Charles}, \citenamefont {Hocker}, \citenamefont {Lacker}, \citenamefont
  {Laplace}, \citenamefont {Le~Diberder}, \citenamefont {Malcles},
  \citenamefont {Ocariz}, \citenamefont {Pivk},\ and\ \citenamefont
  {Roos}}]{Charles:2004jd}%
  \BibitemOpen
  \bibfield  {author} {\bibinfo {author} {\bibfnamefont {J.}~\bibnamefont
  {Charles}}, \bibinfo {author} {\bibfnamefont {A.}~\bibnamefont {Hocker}},
  \bibinfo {author} {\bibfnamefont {H.}~\bibnamefont {Lacker}}, \bibinfo
  {author} {\bibfnamefont {S.}~\bibnamefont {Laplace}}, \bibinfo {author}
  {\bibfnamefont {F.~R.}\ \bibnamefont {Le~Diberder}}, \bibinfo {author}
  {\bibfnamefont {J.}~\bibnamefont {Malcles}}, \bibinfo {author} {\bibfnamefont
  {J.}~\bibnamefont {Ocariz}}, \bibinfo {author} {\bibfnamefont
  {M.}~\bibnamefont {Pivk}},\ and\ \bibinfo {author} {\bibfnamefont
  {L.}~\bibnamefont {Roos}} (\bibinfo {collaboration} {CKMfitter Group}),\
  }\bibfield  {title} {\bibinfo {title} {{CP violation and the CKM matrix:
  Assessing the impact of the asymmetric $B$ factories}},\ }\href
  {https://doi.org/10.1140/epjc/s2005-02169-1} {\bibfield  {journal} {\bibinfo
  {journal} {Eur. Phys. J. C}\ }\textbf {\bibinfo {volume} {41}},\ \bibinfo
  {pages} {1} (\bibinfo {year} {2005})},\ \bibinfo {note} {(updated results and
  plots available at: http://ckmfitter.in2p3.fr)},\ \Eprint
  {https://arxiv.org/abs/hep-ph/0406184} {arXiv:hep-ph/0406184} \BibitemShut
  {NoStop}%
\bibitem [{\citenamefont {Aaij}\ \emph {et~al.}(2020)\citenamefont {Aaij} \emph
  {et~al.}}]{LHCb:2020kho}%
  \BibitemOpen
  \bibfield  {author} {\bibinfo {author} {\bibfnamefont {R.}~\bibnamefont
  {Aaij}} \emph {et~al.} (\bibinfo {collaboration} {LHCb}),\ }\bibfield
  {title} {\bibinfo {title} {{Updated LHCb combination of the CKM angle
  $\gamma$}},\ }\href@noop {} {\  (\bibinfo {year} {2020})},\ \bibinfo {note}
  {{LHCb-CONF-2020-003}}\BibitemShut {NoStop}%
\bibitem [{\citenamefont {Aaij}\ \emph
  {et~al.}(2021{\natexlab{f}})\citenamefont {Aaij} \emph
  {et~al.}}]{LHCb:2021dcr}%
  \BibitemOpen
  \bibfield  {author} {\bibinfo {author} {\bibfnamefont {R.}~\bibnamefont
  {Aaij}} \emph {et~al.} (\bibinfo {collaboration} {LHCb}),\ }\bibfield
  {title} {\bibinfo {title} {{Simultaneous determination of CKM angle $\gamma$
  and charm mixing parameters}},\ }\href@noop {} {\  (\bibinfo {year}
  {2021}{\natexlab{f}})},\ \Eprint {https://arxiv.org/abs/2110.02350}
  {arXiv:2110.02350 [hep-ex]} \BibitemShut {NoStop}%
\bibitem [{\citenamefont {Blanke}\ and\ \citenamefont
  {Buras}(2019)}]{Blanke:2018cya}%
  \BibitemOpen
  \bibfield  {author} {\bibinfo {author} {\bibfnamefont {M.}~\bibnamefont
  {Blanke}}\ and\ \bibinfo {author} {\bibfnamefont {A.~J.}\ \bibnamefont
  {Buras}},\ }\bibfield  {title} {\bibinfo {title} {{Emerging $\Delta M_{d}$
  -anomaly from tree-level determinations of $|V_{cb}|$ and the angle $\gamma
  $}},\ }\href {https://doi.org/10.1140/epjc/s10052-019-6667-x} {\bibfield
  {journal} {\bibinfo  {journal} {Eur. Phys. J. C}\ }\textbf {\bibinfo {volume}
  {79}},\ \bibinfo {pages} {159} (\bibinfo {year} {2019})},\ \Eprint
  {https://arxiv.org/abs/1812.06963} {arXiv:1812.06963 [hep-ph]} \BibitemShut
  {NoStop}%
\bibitem [{\citenamefont {Bona}\ \emph {et~al.}(2006)\citenamefont {Bona} \emph
  {et~al.}}]{UTfit:2006vpt}%
  \BibitemOpen
  \bibfield  {author} {\bibinfo {author} {\bibfnamefont {M.}~\bibnamefont
  {Bona}} \emph {et~al.} (\bibinfo {collaboration} {UTfit}),\ }\bibfield
  {title} {\bibinfo {title} {{The Unitarity Triangle Fit in the Standard Model
  and Hadronic Parameters from Lattice QCD: A Reappraisal after the
  Measurements of $\Delta M_s$ and BR$(B\to \tau \nu_\tau)$}},\ }\href
  {https://doi.org/10.1088/1126-6708/2006/10/081} {\bibfield  {journal}
  {\bibinfo  {journal} {JHEP}\ }\textbf {\bibinfo {volume} {10}},\ \bibinfo
  {pages} {081}},\ \bibinfo {note} {(updated results and plots available at:
  http://www.utfit.org)},\ \Eprint {https://arxiv.org/abs/hep-ph/0606167}
  {arXiv:hep-ph/0606167} \BibitemShut {NoStop}%
\bibitem [{\citenamefont {King}\ \emph
  {et~al.}(2019{\natexlab{b}})\citenamefont {King}, \citenamefont {Kirk},
  \citenamefont {Lenz},\ and\ \citenamefont {Rauh}}]{King:2019rvk}%
  \BibitemOpen
  \bibfield  {author} {\bibinfo {author} {\bibfnamefont {D.}~\bibnamefont
  {King}}, \bibinfo {author} {\bibfnamefont {M.}~\bibnamefont {Kirk}}, \bibinfo
  {author} {\bibfnamefont {A.}~\bibnamefont {Lenz}},\ and\ \bibinfo {author}
  {\bibfnamefont {T.}~\bibnamefont {Rauh}},\ }\bibfield  {title} {\bibinfo
  {title} {{$|V_{cb}|$ and $\gamma$ from $B$-mixing - Addendum to ''$B_s$
  mixing observables and $|V_{td}/V_{ts}|$ from sum rules''}}\ }\href
  {https://doi.org/10.1007/JHEP03(2020)112} {10.1007/JHEP03(2020)112} (\bibinfo
  {year} {2019}{\natexlab{b}}),\ \bibinfo {note} {[Addendum: JHEP 03, 112
  (2020)]},\ \Eprint {https://arxiv.org/abs/1911.07856} {arXiv:1911.07856
  [hep-ph]} \BibitemShut {NoStop}%
\end{thebibliography}%
%%%%%%%%%%%%%%%%%%%%%%%%%%%%%%%%%

\end{document}